\documentclass[twocolumn,aps,prb,preprintnumbers,amsmath,showpacs,amssymb]{revtex4-1}
\usepackage{amsmath}     
\usepackage{amssymb}
\usepackage{amsxtra}
\usepackage{amsfonts}
\usepackage{color}
\usepackage{bm}
\usepackage{dcolumn}
\usepackage{graphicx}

\renewcommand{\vec}[1]{\mathbf{#1}}

\usepackage[active]{srcltx}

\usepackage[usenames,dvipsnames,svgnames,table]{xcolor}

\begin{document}

\preprint{APS/123-QED}

\title{First-principles calculation of the Gilbert damping parameter via
 the linear response formalism with application to magnetic transition-metals and alloys}%

\author{S.~Mankovsky$^1$}
\author{D.~K\"odderitzsch$^1$}
\author{G.~Woltersdorf$^2$}
\author{H.~Ebert$^1$}
\affiliation{$^1$
University of Munich,  
Department of Chemistry, 
Butenandtstrasse 5-13, D-81377 Munich, Germany
}%
\affiliation{$^2$ Department of Physics, Universit{\"a}t Regensburg, 93040 Regensburg, Germany
}%

\date{\today}

\begin{abstract}
A method for the calculations of the Gilbert damping 
parameter $\alpha$ is presented, which based on the linear response formalism, 
has been implemented within the fully relativistic Korringa-Kohn-Rostoker
 band structure method in combination with  the
coherent potential approximation  alloy theory.
To account for thermal displacements of atoms as a 
scattering mechanism, an alloy-analogy model is introduced.
This allows the determination of $\alpha$ for various types of materials,
such as elemental magnetic systems and ordered magnetic compounds at
finite temperature, as well as for disordered magnetic alloys at $T
= 0$~K and above. The effects of spin-orbit coupling, chemical and
temperature induced structural disorder are analyzed.  
Calculations have been performed for the 3$d$ transition-metals
bcc Fe, hcp Co, and fcc Ni, their binary
alloys bcc Fe$_{1-x}$Co$_{x}$,  fcc Ni$_{1-x}$Fe$_x$, 
 fcc Ni$_{1-x}$Co$_x$ and bcc Fe$_{1-x}$V$_{x}$, and for 
$5d$ impurities  in transition-metal alloys. All results are in  satisfying agreement with 
experiment. 
\end{abstract}

\maketitle

\section{\label{sec:level1} Introduction }

During the last decades dynamical magnetic properties have attracted a lot of interest
due to their importance in the development of new devices for spintronics, in
particular, concerning their miniaturization and fast time scale applications. A distinctive
property  of such devices is the magnetization relaxation rate
characterizing the time scale on which a system being deviated from the
equilibrium returns to it,
or how fast the device  can be switched from one state to another. In the
case of dynamics of a uniform magnetization $/vec{M}$ this property is associated
with the Gilbert damping parameter $\tilde G(\vec{M})$  used first in the phenomenological
Landau-Lifshitz (LL) \cite{LL35} and 
Landau-Lifshitz-Gilbert (LLG) theory \cite{Gil04} describing the
magnetization dynamics processes by means of the equation:
%
\begin{eqnarray}
\frac{1}{\gamma}\frac{d\vec{M}}{d\tau}  &= & 
-\vec{M}\times\vec{H}_{\rm eff}
+ \vec{M} \times  \left[ \frac{\tilde{G}(\vec{M})}{\gamma^2 M_s^2}
\frac{d\vec{M}}{d\tau} \right] \;,
\label{LLG}
\end{eqnarray}
%
 where  $ M_s$ is the saturation magnetization, $\gamma$  
the gyromagnetic ratio and
 $\vec{H}_{eff} = - \partial_{\vec{M}} F[\vec{M}(\vec r)]$ being the effective
magnetic field. Sometimes it is more convenient to use 
a dimensionless Gilbert damping parameter $\alpha$ given by  $\alpha 
= \tilde{G}/(\gamma M_s)$ (see, e.g. \cite{GIS07,BTB08,SKB+10}).
Safonov has generalized the Landau-Lifshitz equation by introducing a
tensorial form for the Gilbert damping parameter with the diagonal terms
characterising magnetization dissipation \cite{Saf02}.
Being introduced as a phenomenological parameter, the Gilbert damping 
is normally deduced from experiment. In particular, it can be 
evaluated from  the resonant line width in 
ferromagnetic-resonance (FMR) experiments. 
The difficulty of these measurements consists in the problem 
that there exist several different sources for
the broadening of the line width, which
have been discussed extensively in the literature
\cite{AR55,Jir82,Suh98,HUW02,AM99,KP75,KP72}.
The line width that is observed in ferromagnetic resonance spectra is
usually caused by intrinsic and extrinsic relaxation effects. The
extrinsic contributions are a consequence of spatially fluctuating
magnetic properties due to sample imperfections. Short range 
fluctuations lead to two magnon scattering while long range fluctuations
lead to an inhomogeneous line broadening due a superposition of local
resonances \cite{MTA03}. In order to separate the intrinsic
Gilbert damping from the extrinsic effects it is necessary to measure
the frequency and angular dependence of the ferromagnetic resonance line
width, e.~g. two magnon scattering can be avoided when the magnetization
is aligned along the film normal \cite{AM99} 
(perpendicular configuration). Usually one finds a linear frequency dependence
with a zero frequency offset and one can write $\Delta
H(\omega)=\alpha\frac{\omega}{\gamma}+\Delta H(0)$. When such
measurements are performed over a wide frequency range the slope of  
 $\Delta H$ as a function of frequency can be used to extract the 
 intrinsic Gilbert damping constant. In metallic ferromagnets Gilbert 
 damping is mostly caused by electron magnon scattering. In addition 
 Gilbert-like damping can be caused by eddy currents. The magnitude of 
 the eddy current damping is proportional to  $d^2$, where $d$ is the 
 sample thickness \cite{HUW02}. In sufficiently thin magnetic films ($d\leq 
 10$~nm) the eddy current damping can be neglected \cite{HUW02}. However, for 
 very thin films relaxation mechanisms that occur at the interfaces can 
 also increase and even dominate the damping. Such effects are spin 
 pumping  \cite{TBBH05,BTB11} and the modified electronic structure 
 at the interfaces. In the present work spin pumping and the modified 
 interface electronic structure are not considered and we assume that bulk-like 
 Gilbert damping dominates.

Much understanding of dynamical magnetic properties could in principle
be obtained  from the simulation of these processes utilizing
time-dependent first-principles electronic structure calculations, that
in turn would pave  the way  to developing and optimizing  new materials
for spintronic devices. 
In spite  of the progress  in the development of  time-dependent density
functional theory (TD-DFT) during the last decades \cite{MUN+06}
that allows to study
various  dynamical   processes  in   atoms  and  molecules   from  first
principles, applications to solids are rare.  This is due to a
lack of universally applicable approximations to the 
exchange-correlation kernel  of TD-DFT for solids.
Thus, at the moment, a tractable approach consists in the use of
the classical LLG equations, and  employing  parameters calculated
within a microscopic approach. 
Note however that this approach can fail dealing with ultrafast magnetization
dynamics, which is discussed, for instance, in Refs.\ [\onlinecite{FI11,
  BNF12}], but is not considered in the present work.    

Most of the investigations on the magnetization dissipation have been
carried out within  model studies. 
Here one has to mention, in particular, the so-called  $s$-$d$ or $p$-$d$  exchange model
\cite{HFK67,TFH04,ZL04,FSSA06} based on a separate consideration
of the  localized 'magnetic' $d$-electrons and delocalized $s$- and $p$-electrons
mediating the exchange interactions between localized magnetic
moments and responsible for the magnetization dissipation in the
system. As was pointed out by 
Skadsem \emph{et al.} \cite{STBB07}, the dissipation process in this case can be
treated as an energy pumping out of the $d$-electron subsystem into the
$s$-electron bath followed by its dissipation via spin-flip scattering
processes. 
This model gave a rather transparent qualitative picture for the
magnetization relaxation in diluted alloys, e.g. magnetic
semiconductors such as GaMnAs. However, it fails to give quantitative 
agreement with experiment in the case of itinerant metallic
systems (e.g. 3$d$-metal alloys), where the $d$-states are rather delocalised
and strongly hybridized with the $sp$-electrons.
As a consequence the treatment of \emph{all} valence electrons on the
same footing is needed, which  leads to the requirement of
first-principles  calculations of the Gilbert damping going beyond a model-based
evaluation.  

Various such calculations of the Gilbert damping parameter are already
present in the literature. They usually assume a certain dissipation
mechanism, like Kambersky's breathing Fermi surface  (BFS)
\cite{Kam70,FS06}, or more general torque-correlation models (TCM) 
 \cite{Kam76,GIS07}. 
These models include explicitly the spin-orbit coupling (SOC), 
highlighting its key role in the magnetization
dissipation processes.
However, the latter methods used for electronic structure
calculations cannot take explicitly into account disorder in the system
that in turn is responsible for the aforementioned spin-flip scattering
process. 
Therefore, this has to be simulated by using external
 parameters characterizing the finite lifetime of the electronic states.
This weak point was recently addressed by
Brataas \emph{et al.}\ \cite{BTB08}  who described the Gilbert
 damping by means of scattering theory. This development 
supplied the formal basis for the first parameter-free 
investigations on disordered alloys for which the dominant
 scattering mechanism is potential scattering caused
by chemical disorder \cite{SKB+10}. 

Theoretical investigations of the magnetization dissipation by means of
first-principles calculations of the Gilbert damping parameter already brought 
much understanding of the physical mechanisms responsible for this effect.
First of all,  key roles are played by two effects: the SOC of the atomic species 
contained in the system and scattering on various imperfections, either
impurities or structural defects, phonons, etc.
Accounting for the crucial role of scattering processes responsible for
the energy dissipation, different types of scattering phenomena have to be considered.
 One can distinguish between the ordered-compound or
pure-element systems for which electron-phonon scattering is a very
important mechanism for relaxation, and disordered alloys with
dominating scattering processes resulting from randomly distributed atoms of
different types.  
In the first case, the Gilbert damping behavior is rather different at
low and high temperatures.
At high temperature atomic displacements create random
potentials leading to SOC-induced spin-flip scattering.
At low temperature, where the magnetization dissipation is well
described via the BFS (Breathing Fermi-surface) mechanism
\cite{Kam70,FS06}, the spin-conserving  
electron-phonon scattering is required to bring the electronic subsystem
to the equilibrium at every step of the magnetization rotation, i.e.\ to
reoccupy the modified electronic states.

In this contribution we describe a formalism for the calculation of the 
Gilbert damping
equivalent to that of Brataas \emph{et al.}\ \cite{BTB08}, however,
based on the linear responce theory \cite{EMKK11} as implemented in
fully relativistic multiple scattering based  Korringa-Kohn-Rostoker
(KKR) formalism. It will
be demonstrated that this allows to treat elegantly and efficiently
the temperature dependence of $\alpha$ in pure crystals
as well as disordered alloys.

\medskip
\section{\label{sec:theo} Theoretical approach }

To have direct access to real materials and to obtain a deeper understanding of the
origin of the properties observed experimentally, the phenomenological
Gilbert damping parameter has to be treated on a microscopic level. 
This implies to deal with the electrons responsible for the
energy dissipation in the magnetic dynamical processes.
Thus, one equates the
corresponding expressions for the dissipation rate obtained in the
phenomenological and microscopic approaches  $\dot{E}_{\rm mag} =
\dot{E}_{\rm dis}$.
Although a temporal variation of the magnetization is a required condition
for the energy dissipation to occur,
the Gilbert damping parameter is defined in
the limit $\omega \to 0$ (see e.g., Ref.\ [\onlinecite{STBB07}])  and
therefore can be calculated within the adiabatic
approximation.

In the phenomenological LLG theory the time dependent magnetization
$\vec M(t)$ is described by Eq.\ (\ref{LLG}).
Accordingly, the time derivative of the magnetic energy is given by:
%
%
%
\begin{eqnarray}
\dot{E}_{\rm mag} = \vec{H}_{\rm eff}\cdot\frac{d\vec{M}}{d\tau} 
= \frac{1}{\gamma^2}(\dot{\hat{\vec 
    m}})^{T}[\tilde{G}(\vec{M})\dot{\hat{\vec m}}]
\label{MagnE}
\end{eqnarray}
%
where $\hat{\vec m} = \vec{M}/M_s$ denotes the normalized
magnetization. 
To represent the 
Gilbert damping parameter in terms of a microscopic theory, 
following  Brataas \emph{et al.} \cite{BTB08}, the energy dissipation is
associated with the  
electronic subsystem. The  dissipation rate upon the motion of the
magnetization $\dot{E}_{\rm dis} = \left\langle
  \frac{d\hat{H}}{d\tau}\right\rangle$, is determined by the underlying
Hamiltonian $\hat{H}(\tau)$. Assuming a small deviation of the magnetic moment
from the equilibrium the normalized magnetization $\hat{\vec 
  m}(\tau)$ can be expanded around the equilibrium magnetization $\hat{\vec m}_0$ 
\begin{eqnarray}
\hat{\vec m}(\tau) &=& \hat{\vec m}_0 + \vec{u}(\tau)\; ,
\label{m_expansion}
\end{eqnarray}
%
resulting in the expression for the linearized time dependent Hamiltonian 
for the system brought out of equilibrium:
%
\begin{eqnarray}
\hat{H} &=& \hat{H}_{0}(\hat{\vec m}_0) + \sum_\mu
u_\mu\frac{\partial}{\partial u_\mu} \hat{H}(\hat{\vec{m}}_0) \; .
\label{E_expansion}
\end{eqnarray}
%
Due to the small deviation from the equilibrium, $\dot{E}_{\rm dis}$ can be obtained
within the linear response formalism, leading to the expression \cite{BTB08}:
%
\begin{eqnarray}
\dot{E}_{\rm dis} &=& -\pi\hbar \sum_{ij}\sum_{\mu\nu}
\dot{u}_\mu \dot{u}_\nu
\left\langle  \psi_{i}\bigg|  \frac{\partial \hat{H}}{\partial u_{\mu}} \bigg| \psi_{j}\right\rangle
\left\langle  \psi_{j}\bigg| \frac{\partial \hat{H}}{\partial u_{\nu}}
  \bigg| \psi_i\right\rangle 
\nonumber \\
&&\;\;\;\;\;\;\;\;\;\;\;\;\;\;\;\;\;\;\times \delta(E_F - E_i)\delta(E_F - E_{j})
\; ,
\label{E_dot}
\end{eqnarray}
%
where $ E _F$ is the Fermi energy and the sums run
over all eigenstates  of the system.
As Eq.\ (\ref{E_dot}) characterizes the rate of the energy dissipation
upon transition of the system from the tilted state to the equilibrium, it can be 
identified with the corresponding phenomenological quantity in Eq.\ (\ref{MagnE}),
$\dot{E}_{\rm mag} = \dot{E}_{\rm dis}$. This leads to  
an explicit expression for the 
Gilbert damping tensor $  \tilde G $ 
or equivalently for the damping parameter 
 $\alpha = \tilde G /(\gamma M_s)$ (Ref.\ [\onlinecite{BTB08}]):
%
\begin{eqnarray}
\alpha_{\mu\nu}  &=& -\frac{\hbar \gamma}{\pi M_s} \sum_{ij}\sum_{\mu\nu}
\left\langle  \psi_{i}\bigg|  \frac{\partial \hat{H}}{\partial u_{\mu}}
  \bigg|
  \psi_{j}\right\rangle
\left\langle  \psi_{j}\bigg| \frac{\partial \hat{H}}{\partial u_{\nu}}
  \bigg| \psi_i\right\rangle 
\nonumber \\
&&\;\;\;\;\;\;\;\;\;\;\;\;\;\;\;\;\;\;\times \delta(E_F - E_i)\delta(E_F - E_{j})
\; ,
\label{E_dot_alpha}
\end{eqnarray}
%
where the summation is running over all states at the
Fermi surface $E _F$. 

In full analogy to the problem of electric conductivity \cite{But85}, the sum over
eigenstates $|\psi_i\rangle$ may be expressed in terms of
 the retarded single-particle Green's function 
 $\mbox{Im} G^{+}(E_F) = -\pi
\sum_{i} |\psi_{i}\rangle\langle\psi_{i}|\delta(E_F - E_i)$. 
This
leads for the parameter $\alpha $ to a Kubo-Greenwood-like equation:
%
\begin{eqnarray}
\alpha_{\mu\nu} &=& -\frac{\hbar \gamma}{\pi M_s} 
\mbox{Trace} \nonumber \\
&& \left\langle  \frac{\partial \hat{H}}{\partial u_{\mu}} \mbox{Im}\; G^{+}(E_F)
\frac{\partial \hat{H}}{\partial u_{\nu}}  \mbox{Im}\; G^{+}(E_F)
\right\rangle_{c} \; 
\label{alpha2}
\end{eqnarray}
%
with $\langle ... \rangle_{c}$ indicating a configurational average in
case of a disordered system.

The most reliable way to account for spin-orbit coupling as the source
of Gilbert damping is to evaluate Eq.\ (\ref{alpha2}) using a fully
relativistic Hamiltonian within the framework of local spin density
formalism (LSDA) \cite{Ebe00}: 
%
\begin{eqnarray}
\hat{H} = c\vec{\bm\alpha}\cdot\vec{p} +\beta  m c^2 +
 V(\vec r) + \beta \vec{\bm\sigma}\cdot\hat{\vec m}B(\vec r)
\; .
\label{Hamiltonian}           
\end{eqnarray}
%
Here $\alpha_i$ and $\beta$  are the standard Dirac matrices,
$\vec{\bm\sigma}$ denotes the vector of relativistic Pauli matrices,
 and $\vec p $
 is the relativistic momentum operator \cite{Ros61}.
The functions $V(\vec r)$ and
$\vec B = \vec{\bm\sigma}\cdot\hat{\vec m}B(\vec r)$  are the
spin-averaged and spin-dependent  parts, respectively, 
of the LSDA potential. The spin density $\vec m_s(\vec r)$ as well as
the effective exchange field $\vec{B}(\vec r)$ are assummed to be
collinear within the unit cell and aligned along the $z$-direction in the
equilibrium (i.~e. $\vec{m}_{s,0}(\vec r) =  m_s(\vec r) \hat{\vec 
  m}_{0}= m_s(\vec r) \vec e_{z}$ and
$\vec{B}_0(\vec r) =  B(\vec r) \hat{\vec 
  m}_{0} =  B(\vec r) \vec e_{z}$).   
Tilting of the magnetization direction by the angle $\theta$ 
according to  Eq. (\ref{m_expansion}), i.e. 
$\vec{m}_{s}(\vec r) = m_s(\vec r) \hat{\vec 
  m} = m_s(\vec r) (\sin\theta \cos\phi,\sin\theta \sin\phi,\cos\theta)$ and 
$\vec{B}(\vec r) = B(\vec r) \hat{\vec 
  m} $  
 leads to a perturbation term in the Hamiltonian
%
\begin{eqnarray}
  \Delta V(r) & = & \beta \vec{\bm\sigma}\cdot(\hat{\vec 
  m} - \hat{\vec m}_0)B(\vec r) = \beta
\vec{\bm\sigma}\cdot\vec{u} B(\vec r) \; ,
\end{eqnarray}
with (see Eq. (\ref{E_expansion}))
%
\begin{eqnarray}
\frac{\partial}{\partial u_{\mu}}\hat{H}(\hat{\vec{m}}_0) & = &  \beta
\sigma_{\mu} B(\vec r)  \; .
\end{eqnarray}

 The Green's function $G^{+} $ in Eq.\ (\ref{alpha2})
 can be obtained in a very efficient way by using the spin-polarized  
 relativistic version of multiple
 scattering theory \cite{Ebe00} that allows us to treat magnetic solids: 
%
\begin{eqnarray}
 G^{+}(\vec{r},\vec r',E)& = &
 \sum_{\Lambda \Lambda'}
Z^{n}_{\Lambda}(\vec r,E)
\,
  {\tau}_{ \Lambda  \Lambda'}^{nm}(E)
\,
Z^{m\times}_{\Lambda'}(\vec r',E)
\nonumber \\
&&
\hspace{-9.3ex}
-
\delta_{nm}
\sum_{\Lambda }
\left[
Z^{n}_{\Lambda}(\vec r,E)
\,
J^{n\times}_{\Lambda'}(\vec r',E)
\Theta(r_n'-r_n)
\right.
\nonumber
\\
&&
\left.
  +
J^{n}_{\Lambda}(\vec r,E)
\,
Z^{n\times}_{\Lambda'}(\vec r',E)
\Theta(r_n-r_n')
\right]
.
\label{GreensFunction}
\end{eqnarray}
%
Here $\vec{r},\vec{r}'$ refer to site $n$ and $m$, respectively, where
$Z^{n}_{\Lambda}(\vec r,E)=Z_{\Lambda}(\vec r_n,E) = Z_{\Lambda}(\vec r-\vec R_n
,E) $ is a function centered at site $\vec R_n$.
The four-component wave functions $Z^{n}_{\Lambda}(\vec r,E)$ 
($J^{n}_{\Lambda}(\vec r,E)$) are
regular (irregular)
solutions to the single-site Dirac equation labeled by the
combined quantum numbers $\Lambda$ ($\Lambda = (\kappa,\mu)$), with
$\kappa$ and $\mu$  being the spin-orbit and magnetic quantum numbers
\cite{Ros61}. The superscript $\times$ indicates the left hand
side solution of the Dirac equation.
${\tau}^{nm}_{ \Lambda  \Lambda'} (E)$ is
 the so-called scattering path operator
 that transfers an electronic wave coming in at 
site $ m $ into a wave going out from site $ n $ with
all possible intermediate scattering events accounted for. 

 Using matrix notation with respect to $\Lambda$, this
leads to the following expression for the damping parameter:
\begin{eqnarray}
\alpha_{\mu\mu} =   \frac{g}{\pi\mu_{tot}} \sum_{n } \mbox{ Trace}
\left\langle \underline{T}^{0\mu} \,
 \tilde{\underline{\tau}}^{0n}\,
\underline{T}^{n\mu} \,
 \tilde{\underline{\tau}}^{n0} \right\rangle_{c}
\label{alpha_MST}
\end{eqnarray}
%
with the g-factor $2(1 + {\mu_{orb}}/{\mu_{spin}})$ in terms of
the spin and orbital moments, $\mu_{spin}$ and $\mu_{orb}$,
respectively, the total magnetic moment $\mu_{tot} =
\mu_{spin}+\mu_{orb}$, 
$\tilde{\tau}_{\Lambda\Lambda'}^{0n} =
\frac{1}{2i}(\tau_{\Lambda\Lambda'}^{0n} -
  \tau_{\Lambda'\Lambda}^{0n})$ and with the energy argument $E_F$
omitted.
The matrix elements in Eq. (\ref{alpha_MST})
are identical to those occurring in the context of exchange coupling
\cite{EM09a}:  
%

\begin{eqnarray}
T_{\Lambda'\Lambda}^{n\mu} 
& = & \int d^3r\; Z^{n\times}_{\Lambda'}(\vec{r})\;
\left[\frac{\partial}{\partial u_\mu}\hat{H}(\hat{\vec{m}}_0)\right] Z^{n}_{\Lambda}(\vec{r})
 \nonumber \\
& = & \int d^3r\; Z^{n\times}_{\Lambda'}(\vec{r})\;\left[\beta
\sigma_{\mu}B_{xc}(\vec{r})\right] Z^{n}_{\Lambda}(\vec{r})
\label{matrix-element}
\; .
\end{eqnarray}
%

The expression in Eq.\ (\ref{alpha_MST}) for the Gilbert-damping
parameter $ \alpha $ is essentially equivalent to the one obtained 
within the torque correlation method (see
e.g. Refs.\ [\onlinecite{Kam07,HVT07,GGSM09}]). However, in contrast to the
conventional TCM the electronic structure is here  
represented using the retarded electronic Green function giving the
present  approach much more flexibility.
In particular, it does not rely on a phenomenological relaxation time
parameter.
 
The expression Eq.\ (\ref{alpha_MST})
can be applied straightforwardly to disordered alloys. 
This can be done by describing in a first step 
the underlying electronic structure (for $T=0 $~K)
on the basis of the coherent potential approximation (CPA) alloy theory.
 In the next step the configurational average in Eq. (\ref{alpha_MST})
 is taken following the scheme worked out by Butler \cite{But85}
 when dealing with the electrical conductivity at $T=0 $~K
 or residual resistivity, respectively, of disordered alloys.
This implies in particular that so-called vertex corrections of the type
 $\left\langle  T_{\mu} \mbox{Im} G^{+} T_{\nu} \mbox{Im} G^{+}
\right\rangle_{c} -
\left\langle  T_{\mu} \mbox{Im} G^{+}\right\rangle_{c}
\left\langle  T_{\nu} \mbox{Im} G^{+}\right\rangle_{c}
$
 that account for scattering-in processes in the language of the 
Boltzmann transport formalism are properly accounted for.

One has to note that the factor  $\frac{g}{\mu_{tot}}$  in  Eq.\
(\ref{alpha_MST}) is separated from the configurational average
$\left\langle  ...\right\rangle_{c}$, although both values, $g$
and ${\mu_{tot}}$, have to represent the average per unit cell doing the
calculations for compounds and alloys.
This approximation is rather reasonable  in the case
of compounds or alloys where the properties of the elements of
the system are similar (e.g. 3$d$-element alloys), but can be questionable
in the case of systems containing elements exhibiting significant
differences (3$d$-5$d$-, 3$d$-4$f$-compounds, etc), or in the case of
non-uniform systems as discussed by Nibarger et al \cite{NLCS03}.

Thermal vibrations as a source of electron scattering can in principle
be accounted for by a generalization of Eqs.\  (\ref{alpha2}) --
(\ref{matrix-element}) to finite temperatures and by including the
electron-phonon self-energy $\Sigma_{el-ph}$ when calculating the Green's
function $G^+$. Here we restrict our consideration to elastic scattering
processes by using a quasi-static representation of the thermal
displacements of the atoms from their equilibrium positions. 
The atom displaced from the equilibrium position in the lattice
results in a corresponding variation $\Delta \underline{t}^n =
\underline{t}^n -  \underline{t}_0^n $ of the single-site scattering
matrix in the global frame of reference \cite{PZDS97, Lod76}. 
A single-site scattering matrix $\underline{t}^n$ 
(the underline denotes a matrix in an angular momentum representation $\Lambda$)
for the atom $n$ displaced  by the value $\vec{s}^n_{\nu}$
from the equilibrium position in the lattice 
can be obtained using the transformation matrices \cite{SBZD87, PZDS97}
\begin{eqnarray}
U^n_{LL'}(\vec{s}_{\nu},E) & = & 4\pi \sum_{L''}
i^{l''+l-l'} \nonumber\\
 & \times & C_{LL'L''}j_{l''}(s^n_{\nu}\sqrt{E})Y_{L''}(\hat{\vec 
  s}^n_{\nu})  
\label{U_displacement}
\; .
\end{eqnarray}
%
Here $m_e$ is the electron mass, $j_{l}$ a spherical Bessel function,
$C_{LL'L''}$  stands for the Gaunt coefficients, and a non-relativistic
angular momentum representation with $L = (l,m_l)$ has been
used. Performing a Clebsch-Gordon transformation for the transformation
matrix $U^n_{LL'}$ to the relativistic $\Lambda$ representation, the $t$
matrix $\underline{t}^n$ for the shifted atom can be obtained from the
non-shifted one $\underline{t}^n_0$ from the expression 

\begin{eqnarray}
\underline{t}^n_{\nu} & = & (\underline{U}^n_{\nu})^{-1}\underline{t}_0^n \underline{U}^n_{\nu}
\label{t_displaced}
\;.
\end{eqnarray}

Treating for a discrete set of displacements ${{\vec s}^n_{\nu}}$ each
displacement as an alloy component, we introduce an
alloy-analogy model to average over the set ${{\vec s}^n_{\nu}}$
that is chosen to reproduce the thermal root mean square average displacement
$\sqrt{\langle u^2\rangle_T}$  for a given temperature $T $. 
This in turn may be set according to ${\langle u^2\rangle_T} =
\frac{1}{4}\frac{3h^2}{\pi^2mk\Theta_D}[\frac{\Phi(\Theta_D/T)}{\Theta_D/T}
+\frac{1}{4}]$ with 
$\Phi(\Theta_D/T)$ the Debye function,
$h$ the Planck constant, $k$ the Boltzmann constant and
$\Theta_D$ the Debye temperature \cite{GMMP83}.
Ignoring the zero temperature term $1/4$ and
assuming a frozen potential for
the atoms, the situation can be dealt with in full
analogy to the treatment of disordered alloys on the basis of the CPA
(see above).

For small displacements the transformation Eq. (\ref{U_displacement}) 
can be expanded with respect to ${{\vec s}^n_{\nu}}$ 
(see Ref.\ [\onlinecite{SBZD87}]) resulting in a linear 
dependence on ${{\vec s}^n_{\nu}}$ for non-vanishing 
contributions 
with $\Delta l = |l - l'| = \pm 1$. This leads, in particular,
in the presence of atomic displacements
for transition-metals (TM), for which  an angular momentum
cut-off of $l_{\mbox{\tiny  max}} = 2$ in the KKR multiple scattering expansion 
is in general sufficient for an
undistorted lattice,  to an angular momentum expansion 
up to at least $l_{\mbox{\tiny  max}} = 3$.
However, this is correct only 
under the assumption of very small displacements
allowing linearisation of the transformation $U$ with respect to the 
displacement amplitude $s$. Thus, since the temperature increase leads
to a monotonous increase of $s$, the cut-off $l_{\mbox{\tiny  max}}$
should also be increased.

\section{Model calculations}

In the following we present results of calculations for which single
parameters have artificially been manipulated in the first-principles
calculations in order to systematically reveal their role for the
Gilbert-damping. This approach is  used to disentangle competing 
influences on the Gilbert-damping parameter. 

\subsection{Vertex corrections}

\begin{figure}
  \begin{center}
 \includegraphics[scale=0.35]{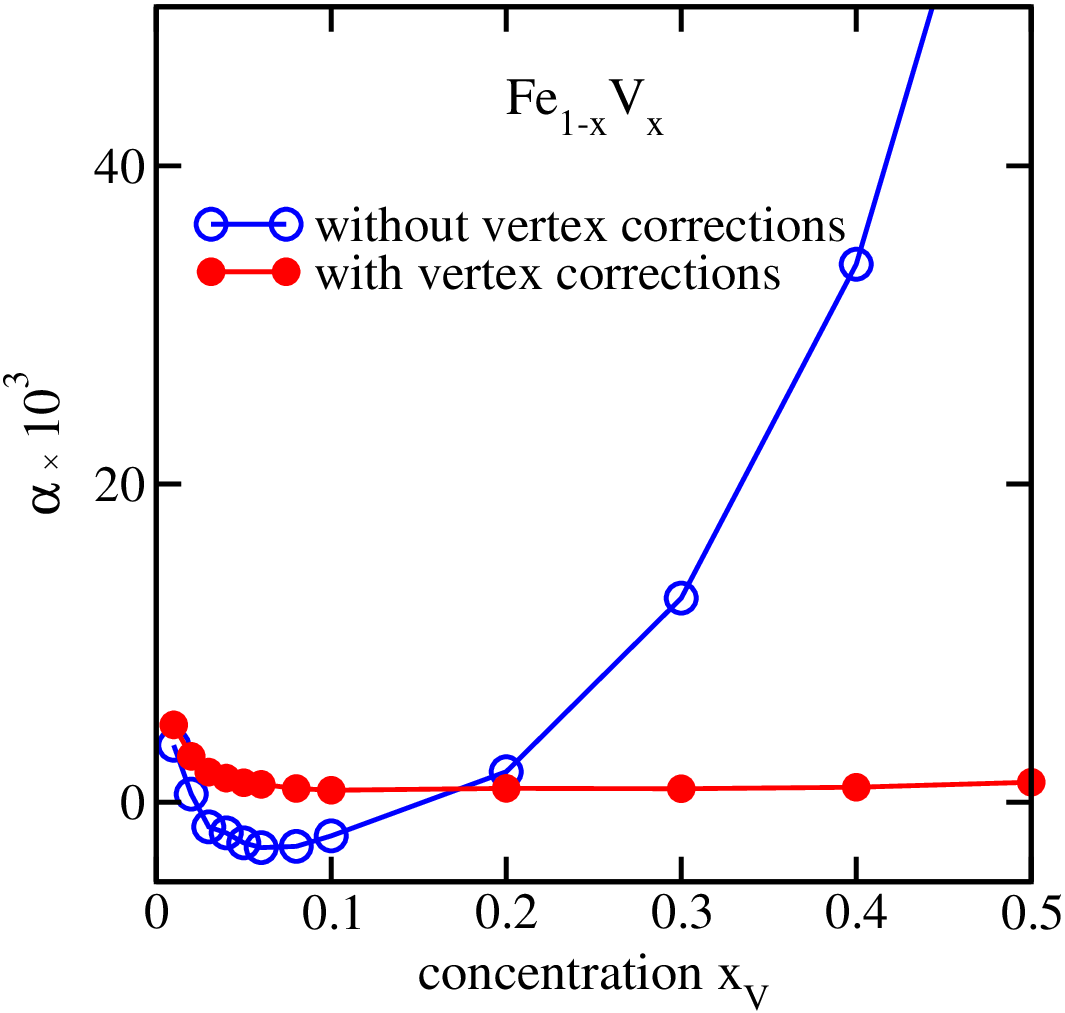} \;\; (a) \\
 \includegraphics[scale=0.35]{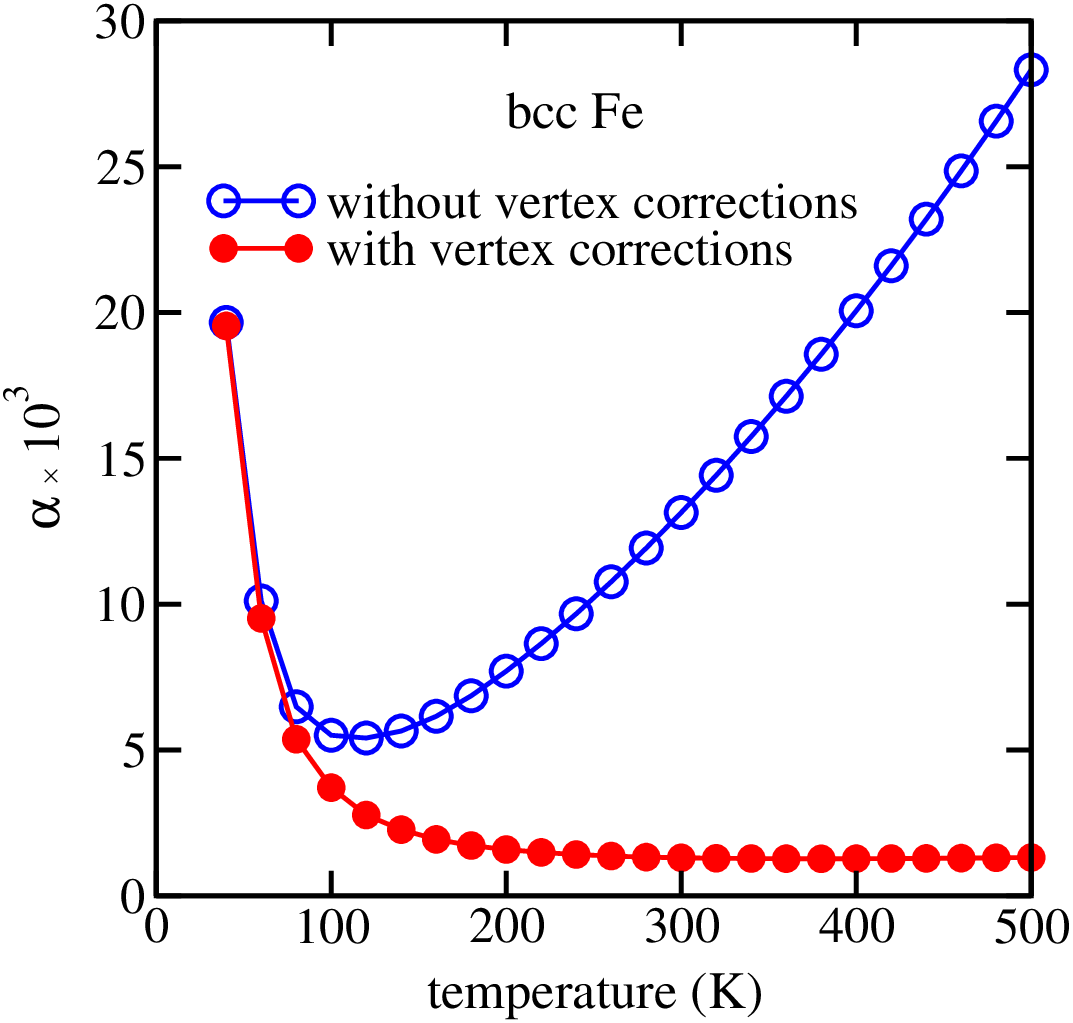}\;\; (b)
  \end{center}
  \vskip-5mm
  \caption{\label{Fig-vertex}
The Gilbert damping parameter for (a) bcc Fe$_{1-x}$V$_x$  ($T = 0$~K) as a function
    of V concentration and (b) for bcc-Fe as a function of
    temperature. Full (open) symbols give results with (without) the
    vertex corrections.} 
  \vskip-6mm  
\end{figure}

The impact of vertex corrections is shown in Fig. \ref{Fig-vertex} for two
different cases: Fig. \ref{Fig-vertex}(a) represents the Gilbert damping
parameter for an Fe$_{1-x}$V$_x$ disordered alloy as a function of 
concentration, while Fig. \ref{Fig-vertex}(b)  gives the corresponding
value for pure Fe in the presence of temperature induced disorder and
plotted as a function of temperature. Both figures show results
calculated with and without vertex corrections allowing for comparison. 
First of all, a significant effect of the vertex corrections is
noticeable in both cases, although the variation depends 
on increasing concentration of V in the binary Fe$_{1-x}$V$_x$
alloy and the temperature in the case of pure Fe, respectively. 
Some differences in their behavior can be explained by the 
differences of the systems under consideration. Dealing with temperature
effects via the alloy analogy model, the system
is considered as an effective alloy consisting of a fixed number of
components characterizing different types of displacements. Thus, in
this case the temperature effect is associated with the increase of disorder 
in the system caused only by the increase of
the displacement amplitude, or, in other words -- with the strength of
scattering potential experienced by the electrons represented by   
$\underline{t}^n(T) - \underline{t}_0^n$. In the case of a random alloy the $A_{1-x}B_x$
variation of the scattering potential, as well as the difference
$\underline{t}^n_A - \underline{t}^n_B$, upon changing the
concentrations is less pronounced for small amounts of impurities $B$
and the concentration dependence is determined by the amount of scatterers
of different types. However, when the concentration of impurities
increases, the potentials of the  
components are also modified (this is reflected, e.g., in the shift of
electronic states with respect to the Fermi level, that will be discussed
below) and this can lead to a change of the concentration dependence of the
vertex corrections. An important issue which one has to stress that
neglect of the vertex corrections may lead to the unphysical result,
$\alpha < 0$, as is shown in Fig. \ref{Fig-vertex}(a). In terms of
the Boltzmann transport formalism, this is because of the neglect of the
scattering-in term \cite{BS84} leading obviously to an incomplete 
description of the energy transfer processes.

\subsection{\label{sec:metall} Influence of spin-orbit coupling }

\begin{figure}
  \begin{center}
 \includegraphics[scale=0.35]{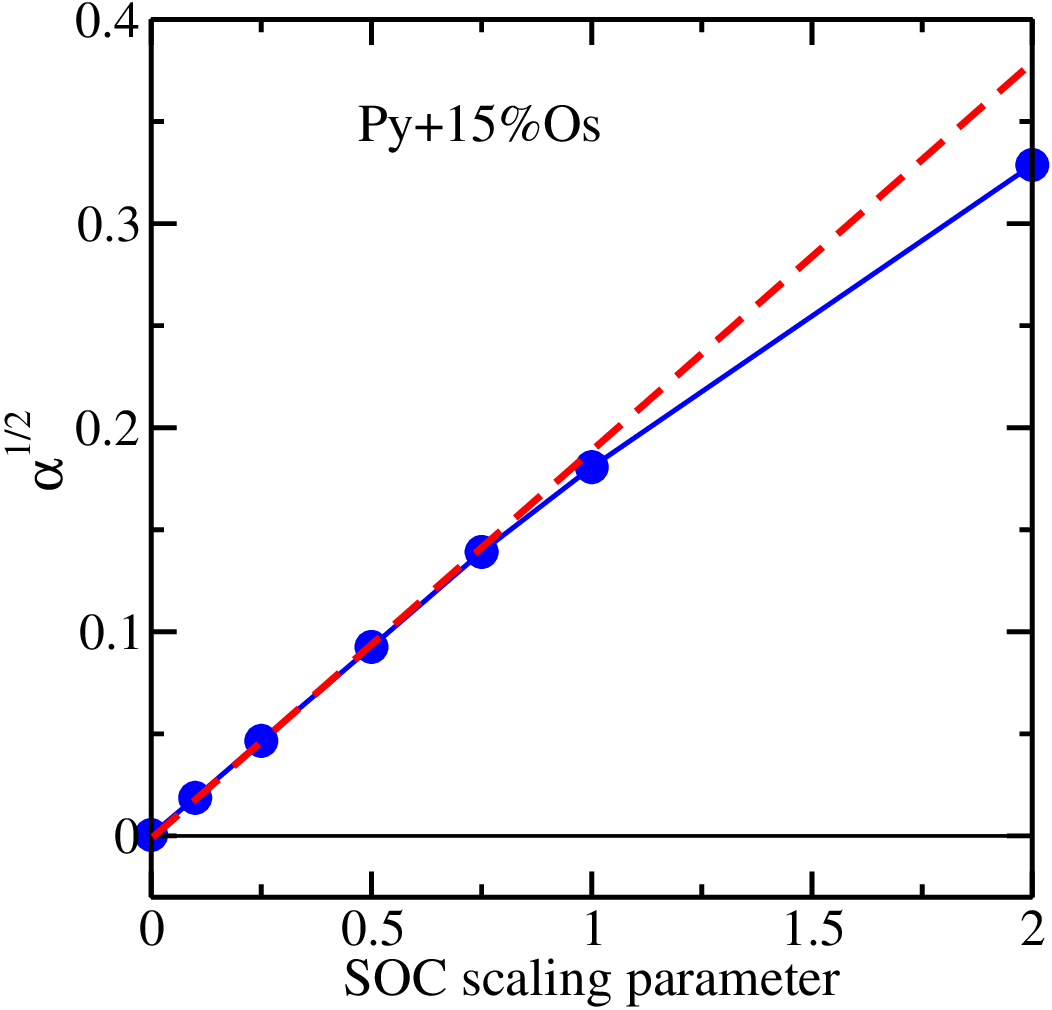}
  \end{center}
  \caption{\label{Fig-SOC_scale}  The   Gilbert  damping  parameter  for
    Py+15\%Os  as a  function  of the  scaling  parameter of  spin-orbit
    coupling applied to all  atoms contained in the alloy.  Red dashed
      line in plot -- linear fit. The values $0$ and $1$ for the SOC scaling parameter correspond to the scalar-relativistic Schr\"odinger-like and fully relativistic Dirac equations, respectively. }
\end{figure}

As was already discussed above, the spin-orbit coupling
for the electrons of the atoms composing the system 
is the main driving force for the magnetization relaxation,
resulting in the energy transfer from the magnetic subsystem to the
crystal lattice. Thus, the Gilbert damping parameter should approach zero
upon decreasing the SOC in the system. 
Fig. \ref{Fig-SOC_scale} shows  the results for  Py+15\%Os, where
$\sqrt{\alpha}$ is plotted as a function of the scaling parameter of the
spin-orbit coupling \cite{EFVG96} applied to all atoms in the alloy. As one can see,
$\sqrt{\alpha}$ has  a nearly linear dependence on SOC
implying that $\alpha$ varies
in second order in the strength of the spin-orbit coupling  \cite{Kam84}.
%

\section{\label{sec:results} Results and discussions}

\subsection{\label{sec:results-TM} 3$d$ transition-metals}

We have mentioned above the crucial role of scattering processes for the energy
dissipation in magnetization  dynamic processes. In pure metals, in
the absence of any impurity, the electron-phonon scattering mechanism
is of great importance, although it plays a different role in the
low- and high-temperature regimes. 
This was demonstrated by Ebert \emph{et al.} \cite{EMKK11} using the alloy
analogy approach, as well as by Liu \cite{LSYK11} \emph{et al.}
using the 'frozen thermal lattice disorder' approach.
In fact both approaches are based on the quasi-static treatment of
thermal displacements. However, while the average is taken by the CPA
within the alloy analogy model the latter requires a sequence of
super-cell calculations for this purpose. 

\begin{figure}
  \begin{center}
 \includegraphics[scale=0.35]{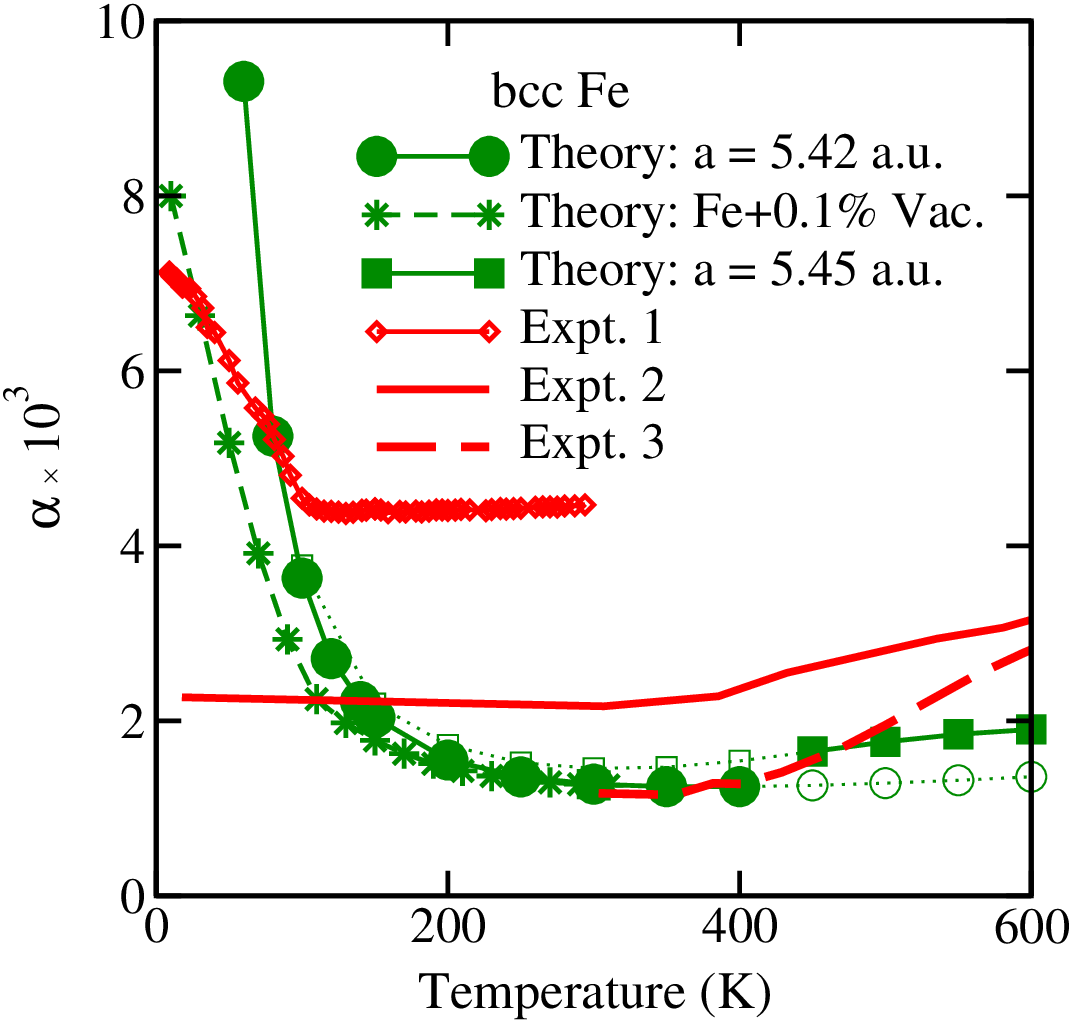}\,(a) \\
 \includegraphics[scale=0.35]{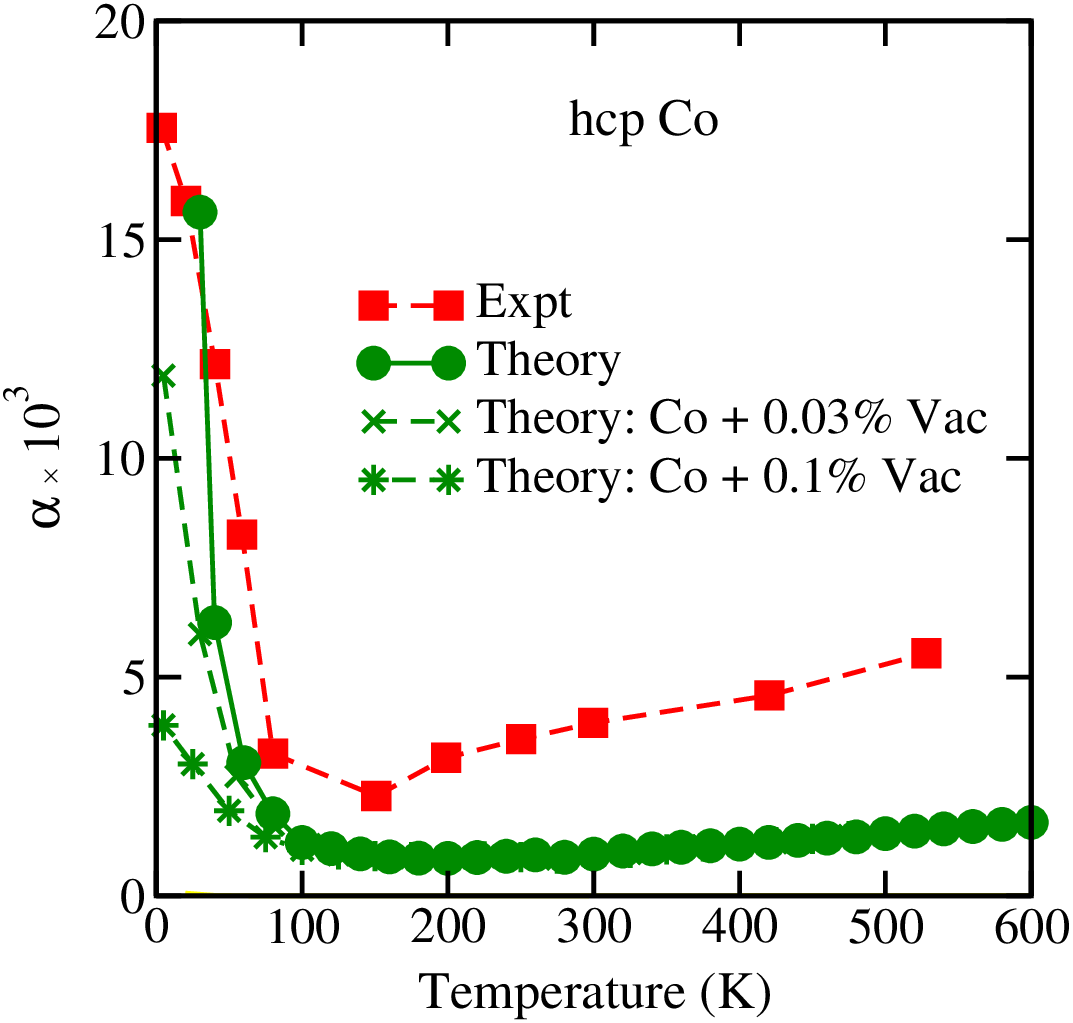}\,(b)\\
 \includegraphics[scale=0.35]{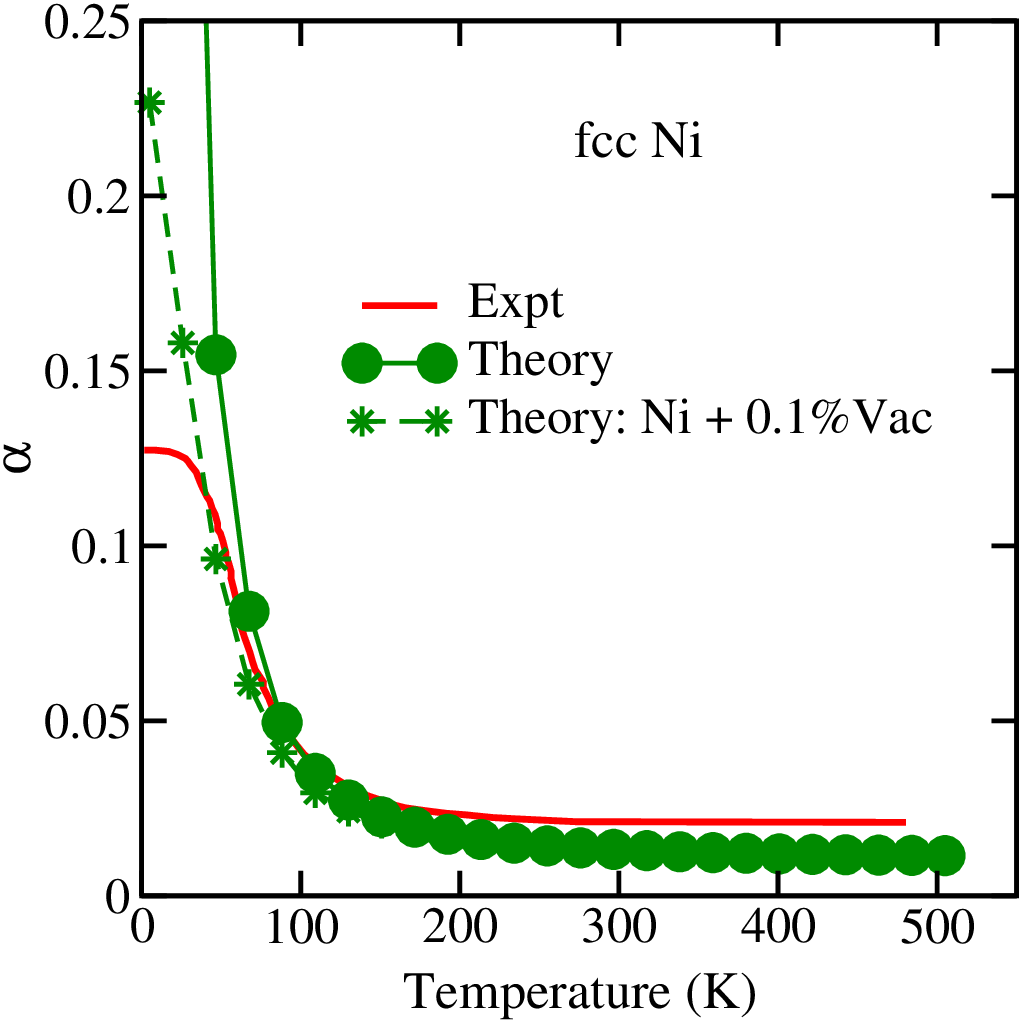}\,(c) 
  \end{center}
  \caption{  \label{Fig-NiFeCo}  Temperature  variation of  the  Gilbert
    damping parameter of pure systems. Comparison of 
    theoretical results with
    experiment: (a) bcc-Fe:
    circles and squares show the results for ideal bcc Fe for two
    lattice parameters, $a = 5.42$~a.u. and $a = 5.45$~a.u.;
    stars show theoretical results for bcc Fe ($a = 5.42$~a.u.) with 0.1\% of vacancies
    (Expt.~1 -  Ref.\ [\onlinecite{Wol11}],   Expt.~2 - Ref.\  [\onlinecite{BL74}],
    Expt.~3 - Ref.\ [\onlinecite{HF70}]); (b) hcp-Co: circles show theoretical
    results for ideal hcp Co, stars - for Co with 0.03\% of vacancies,
    and 'pluses' - for Co with 0.1\% of vacancies (Expt.\ Ref.\ [\onlinecite{BL74}]); and  (c) fcc-Ni (Expt. Ref. [\onlinecite{BL74}]).}
 \end{figure}

As a first example bcc Fe is considered here. The calculations have been
performed accounting for the temperature induced atomic displacements
from their equilibrium positions, according to the alloy analogy scheme described in
section \ref{sec:theo}. This leads, even for pure systems, to a scattering process
and in this way to a finite value for $ \alpha $ (see Fig.  \ref{Fig-NiFeCo}(a)).
One can see that the experimental results available in the literature are rather different,
depending on the conditions of the experiment. In particular, the
experimental results  Expt.~2 (Ref.\ [\onlinecite{BL74}]) and Expt.~3 (Ref.\ [\onlinecite{HF70}]) correspond
to bulk while the measurements Expt.~1 (Ref.\ [\onlinecite{Wol11}])
 have been done for an ultrathin film with $2.3$ nm thickness. 
The Gilbert damping constant obtained within the present calculations for  bcc Fe 
(circles, $a = 5.44$ a.u.) is compared  in Fig.  \ref{Fig-NiFeCo}(a) with
the experiment exhibiting rather good agreement at the temperature
above 100 K despite a certain underestimation. 
One can also see a rather pronounced  increase of the Gilbert damping
observed in the experiment above 400 K (Fig. \ref{Fig-NiFeCo}(a), Expt.~2
and Expt.~3), while the theoretical value shows only little temperature
dependent behavior. Nevertheless, the increase of the
Gilbert damping with  temperature becomes more pronounced when the
temperature induced lattice expansion is taken into account, that can be
seen from the results obtained for $a = 5.45$ a.u. (squares). Thick lines
are used to stress the temperature regions for which corresponding
lattice parameters are more appropriate.  
At low temperatures, below $100$ K, the calculated Gilbert damping
parameter goes up when the temperature decreases, that was observed only in
the recent experiment \cite{Wol11}.
This behavior is commonly denoted as a transition 
from low-temperature conductivity-like to high-temperature resistivity-like behavior
reflecting the dominance of intra- and inter-band
transitions, respectively \cite{GIS07}.
The latter are related to the increase of the smearing of electron energy
bands caused by the increase of scattering events with temperature.
Note that even a small amount of impurities  reduces strongly 
the conductivity-like behavior \cite{BL74,EMKK11},  
leading to the more pronounced effect of impurity-scattering processes
 due to the increase of scattering events caused by chemical disorder.
Large discrepancies between the latter experimental data
\cite{Wol11}  and theoretical results of the $\alpha$ calculations for
bcc Fe are related to the very small thickness of the film investigated
experimentally, that leads to an increase of spin-transfer channels for
magnetization dissipation as was discussed above.  

Results for the temperature dependent Gilbert-damping parameter $\alpha$
for  hcp Co are presented in Fig. \ref{Fig-NiFeCo}(b) which shows,
despite certain underestimation, a
reasonable agreement with the experimental results \cite{BL74}. The
general trends at low and high temperatures are similar to those seen in
Fe.

The results for pure Ni are given in
Fig.\  \ref{Fig-NiFeCo}(c)  that  show in full accordance 
with experiment a rapid decrease 
of $\alpha $ with increasing temperature until 
a regime with a weak variation of $\alpha$ with $T$
is reached. 

\begin{figure}[b]
  \begin{center}
 \includegraphics[scale=0.35]{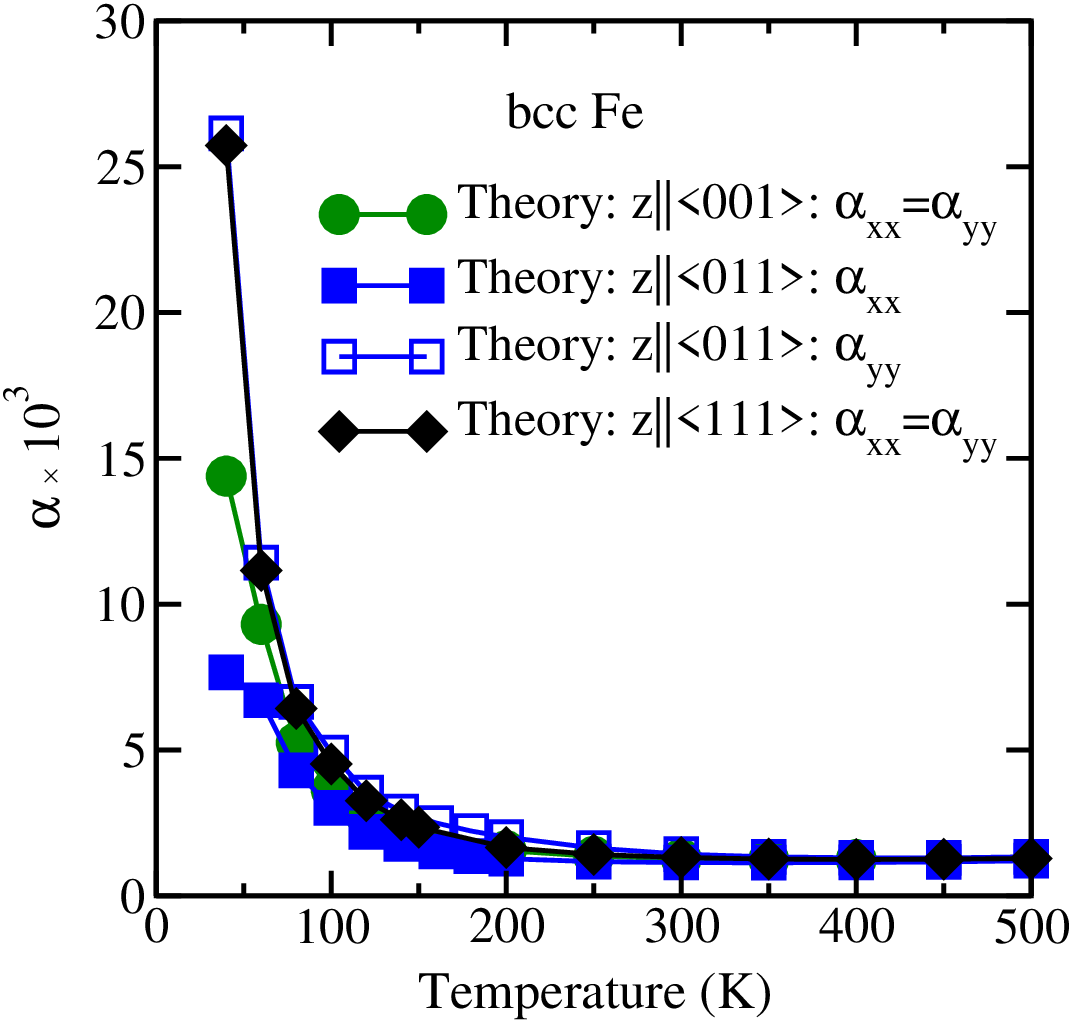}
  \end{center}
  \vskip-5mm
  \caption{ \label{Fig-Fe}
Temperature variation of the $\alpha_{xx}$ and $\alpha_{yy}$ components
of the Gilbert damping tensor of bcc Fe with the magnetization direction taken
along different crystallographic directions:
 $\vec{M} = \hat{z}|\vec{M}| \, \| \langle 001\rangle$ (circles),
 $\vec{M}  \| \langle 011 \rangle$ (squares),
 $\vec{M}  \| \langle 111 \rangle$ (diamonds). }
  \vskip-6mm
 \end{figure}

Note that in the discussions above we have treated  $\alpha$ 
as a scalar instead of a tensorial  quantity
 ignoring a possible anisotropy of the damping processes. 
This approximation is reasonable for the systems considered above with 
the magnetization directions along a three-  or fourfold symmetry
axis (see, e.g., the discussions in Ref.\ [\onlinecite{SF05,GSS+10}]). 
For a more detailed discussion of this issue 
  the anisotropy of the  Gilbert damping tensor
$\underline{\underline{\alpha}}(\vec{M})$
has been investigated for  bcc Fe. To demonstrate  the dependence of
$\underline{\underline{\alpha}}$ on the magnetization direction $\vec{M}$, 
the calculations have been performed for $\vec{M} = \hat{z}|\vec{M}|$
 with  the $\hat{z}$ axis taken along the   $\langle 001 \rangle$,  $\langle 111 \rangle$  and $\langle 011 \rangle$
crystallographic directions. Fig.\ \ref{Fig-Fe} presents the temperature
dependence of the diagonal elements $\alpha_{xx}$ and $\alpha_{yy}$.
As to be expected for symmetry reasons, $\alpha_{xx}$ differs from 
$\alpha_{yy}$ only in the
case of $\hat{z} \| \langle 011 \rangle$. 
One can see that the anisotropic behavior of the Gilbert damping 
 is  pronounced at low temperatures. With an increase of the 
 temperature  the anisotropy nearly disappears, because of the 
 smearing of the energy bands caused by thermal vibrations \cite{GSS+10}. 
 A similar behavior is caused  by 
 impurities with a random distribution, as was observed 
 for example for the  Fe$_{0.95}$Si$_{0.05}$ alloy system. 
The calculations of the diagonal elements $\alpha_{xx}$ and
$\alpha_{yy}$ for two different magnetization directions along
$\langle 001 \rangle$ and $\langle 011 \rangle $ axes give $\alpha_{xx}=\alpha_{yy}=0.00123$  in the first
case and $\alpha_{xx}=0.00123$ and $\alpha_{yy}=0.00127$ in the second,
i.e. the damping is nearly isotropic. 
 
 The damping parameter $\alpha$ increases very rapidly with decreasing temperature 
in the low temperature regime ($T \leq 100$~K) for all pure
ferromagnetic $3d$ metals, Fe, Co, and Ni (see Fig. \ref{Fig-NiFeCo}),
leading to a significant
discrepancy between theoretical and experimental results in this regime. 
The observed discrepancy between  theory and
experiment can be related to the exact limit $\omega = 0$ taken in the
expression for the Gilbert damping parameter. 
Korenmann and Prange \cite{KP72} have analyzed the magnon damping in the
limit of small wave vector of magnons $\vec{q} \to 0$, assuming indirect
transitions in the electron subsystem and taking into account the finite
lifetime $\tau$ of the Bloch states due to electron-phonon
scattering. They discuss the limiting cases of low and high temperatures
showing the analogy of the present problem with the problem of extreme
cases for the conductivity leading to the normal and anomalous skin
effect. On the basis of their result, the authors point out that the
expression for the Gilbert damping obtained by Kambersky \cite{Kam70}, with
$\alpha \sim \tau$ is correct in the limit of small lifetime
(i.e. $qv_F\tau \ll 1$, in their model consideration, where $q$ is a
magnon wave vector and $v_F$ is a Fermi velocity of the electron). In
the low-temperature limit the lifetime $\tau$ increases with decreasing $T$ 
and one has to
use the expression corresponding to the 'anomalous' skin effect for the
conductivity, i.e. $\alpha \sim \tan^{-1}(qv_F\tau)/qv_F$, leading to a
saturation of $\alpha$ upon the increase of $\tau$.

Another possible reason for the low-temperature behavior of the Gilbert damping
observed experimentally can be structural defects present in the
material. To simulate this effect, calculations have been performed
for fcc Ni and bcc Fe with 0.1\% of vacancies and for hcp Co with 0.1\%
and  0.03\% of vacancies. Fig. \ref{Fig-NiFeCo}(a)-(c) shows
 the corresponding  temperature dependence of the Gilbert damping
parameter  approaching  a  finite   value  for  $T\rightarrow  0$.   The
remaining difference in the  $T$-dependent behavior can be attributed to
the  non-linear  dependence  of  the  scattering cross  section  at  low
temperatures as is discussed  in the literature for transport properties
of metals and is not accounted for within the present
approximation.

\subsection{\label{sec:results-alloys} 3$d$ Transition-metal alloys}

As is mentioned above, the use 
of the linear response formalism within multiple scattering
theory for the electronic structure calculations allows us to
 perform the necessary configurational averaging in a very
 efficient way avoiding supercell calculations
 and  to study with moderate effort the 
 influence of varying  alloy composition on $ \alpha $. 
The corresponding approach has been applied to the 
ferromagnetic 3$d$-transition-metal alloy systems
 bcc Fe$_{1-x}$Co$_{x}$,  fcc Ni$_{1-x}$Fe$_x$, 
 fcc Ni$_{1-x}$Co$_x$ and bcc Fe$_{1-x}$V$_{x}$.

\begin{figure}
  \begin{center}
 \includegraphics[scale=0.35]{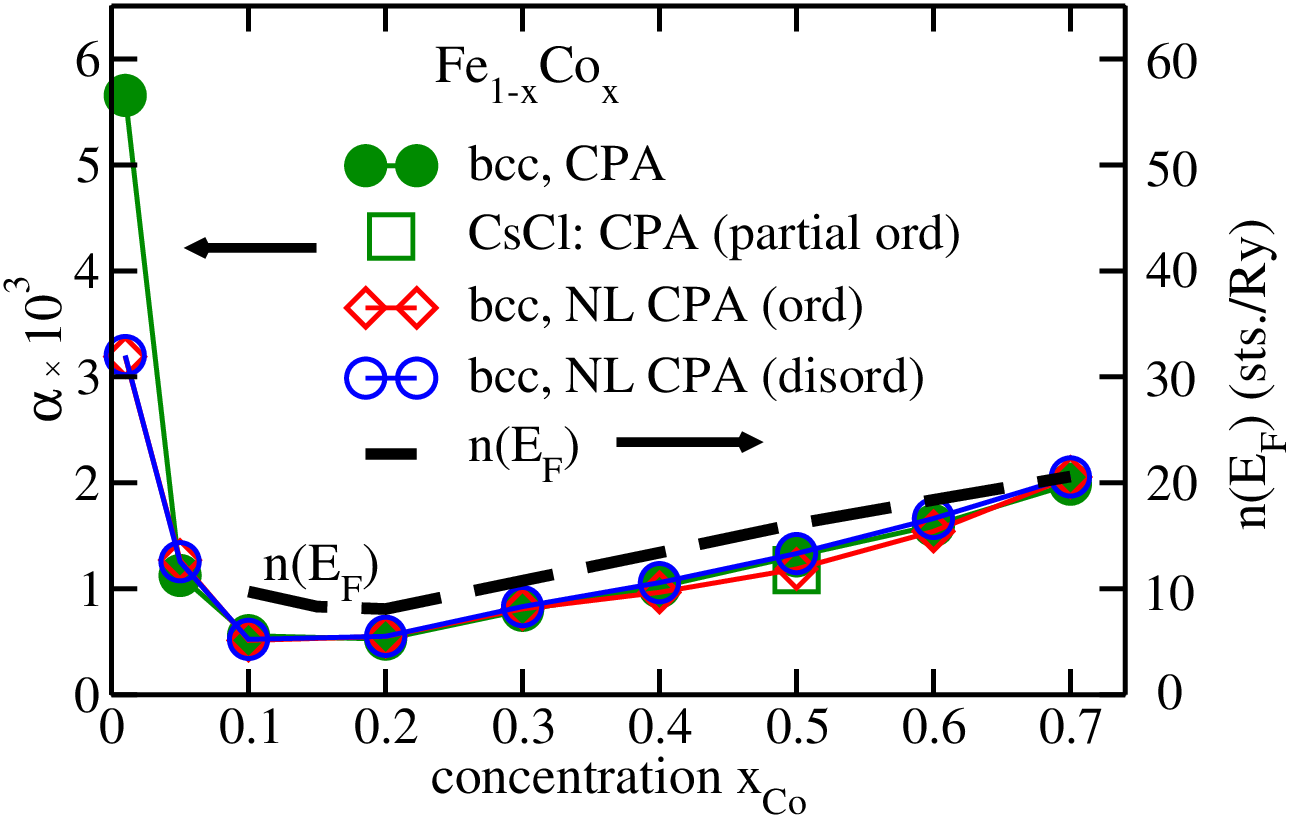}\;(a)
 \includegraphics[scale=0.35]{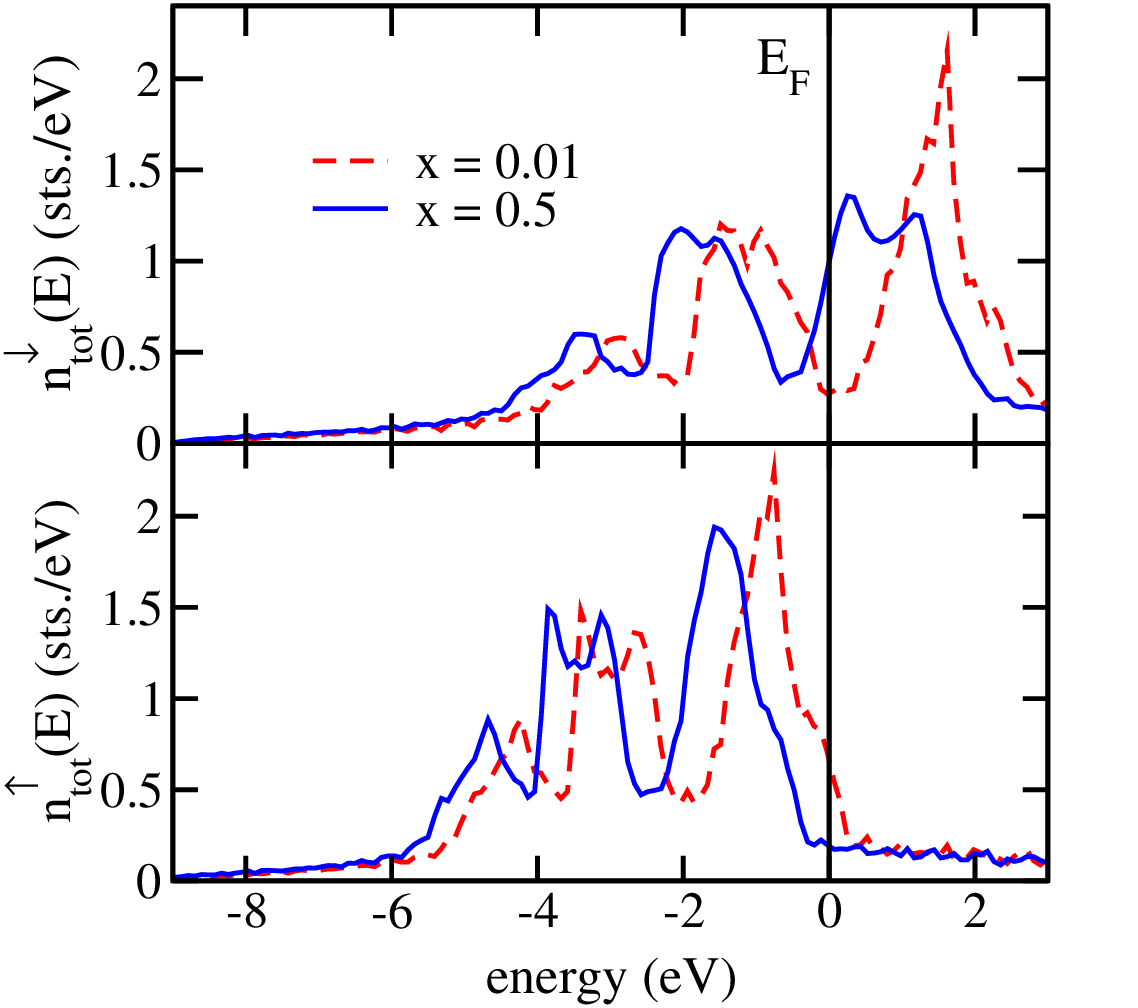}\;\; \;\;\;\;(b)
  \end{center}
  \caption{\label{Fig-FeCo_BSF}
(a) Theoretical results for the Gilbert damping parameter of bcc
Fe$_{1-x}$Co$_{x}$ as a function of Co concentration:  CPA results
for the bcc structure (full circles) describing the random alloy system, 
results for the partially ordered system (opened square) for  $x = 0.5$
(i.e. for Fe$_{1-x}$Co$_{x}$ alloy with CsCl structure and alloy
components randomly distributed in two sublattices in the following
proportions: (Fe$_{0.9}$Co$_{0.1}$)(Fe$_{0.1}$Co$_{0.9}$),
 the NL-CPA results for random alloy with bcc
structure (opened circles) and the NL-CPA results for the the system with
short-range order within the first-neighbor shell (opened diamonds).
The dashed line represents the DOS at the Fermi energy, $E_F$, as a function
of Co concentration.
 (b) spin resolved DOS for bcc  Fe$_{1-x}$Co$_{x}$ for
    $x = 0.01$ (dashed line) and  $x = 0.5$ (solid line). }
\end{figure}

\begin{figure}
  \begin{center}
 \includegraphics[scale=0.30]{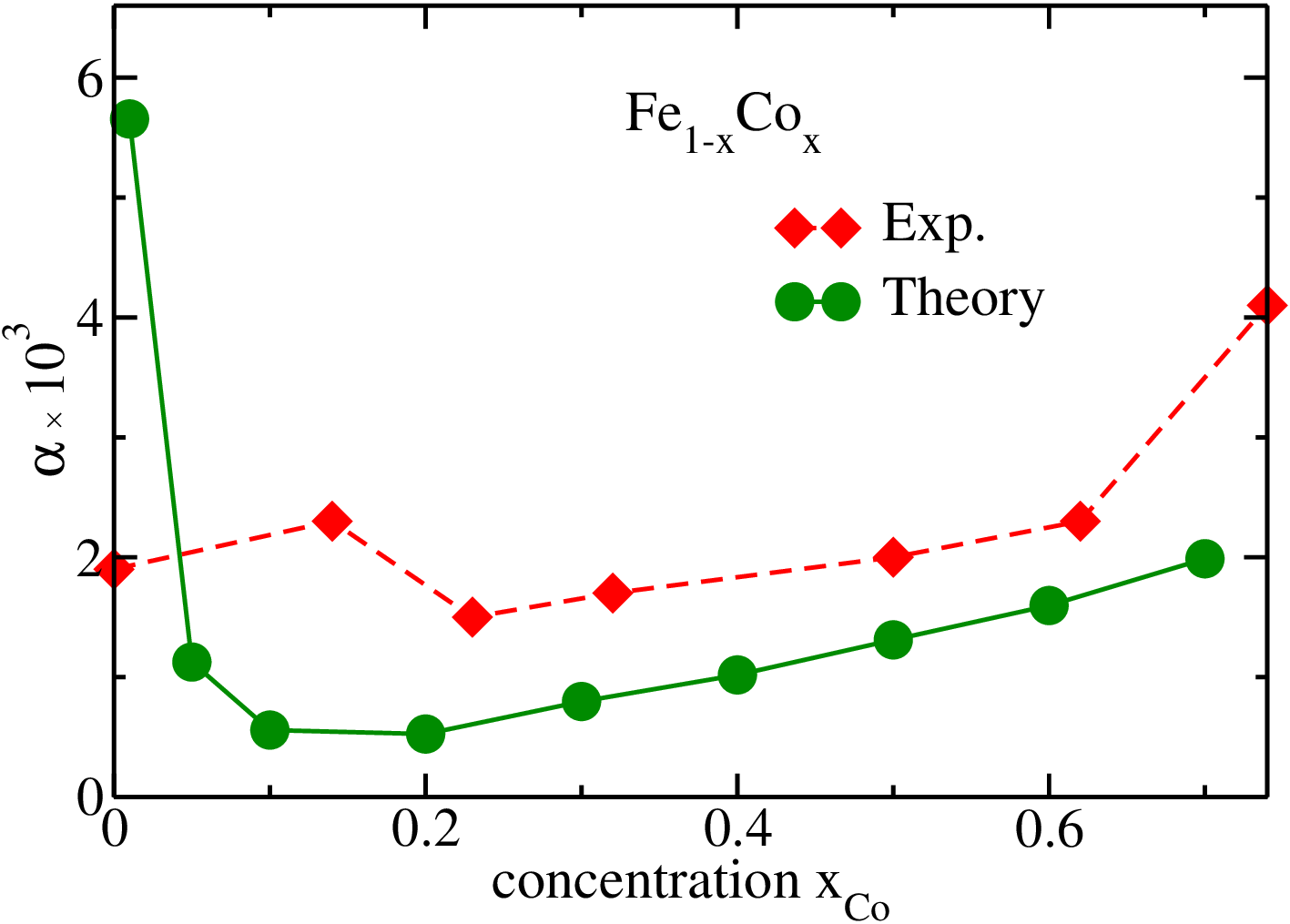}\;(a)
 \includegraphics[scale=0.30]{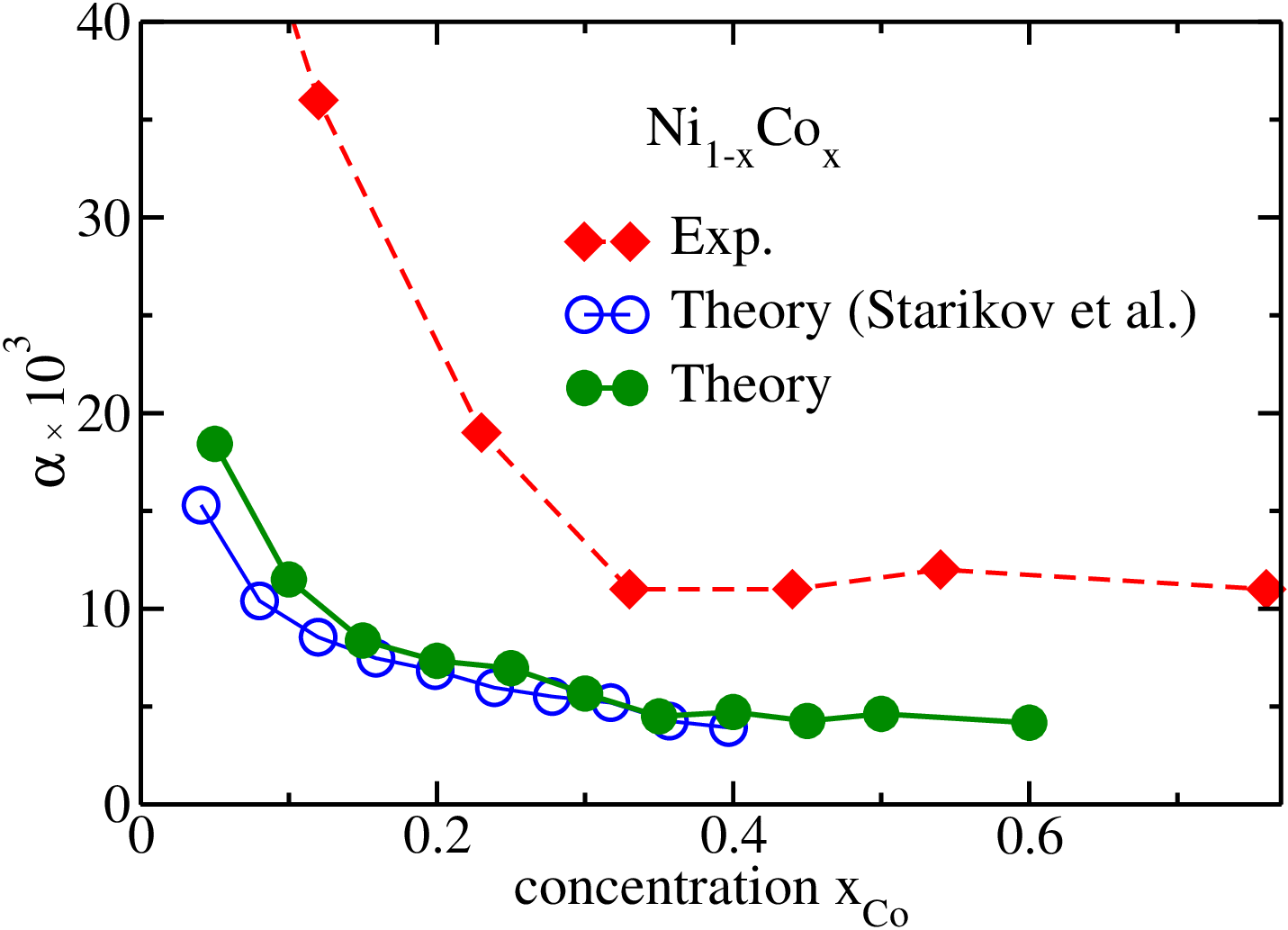}\;(b)
 \includegraphics[scale=0.30]{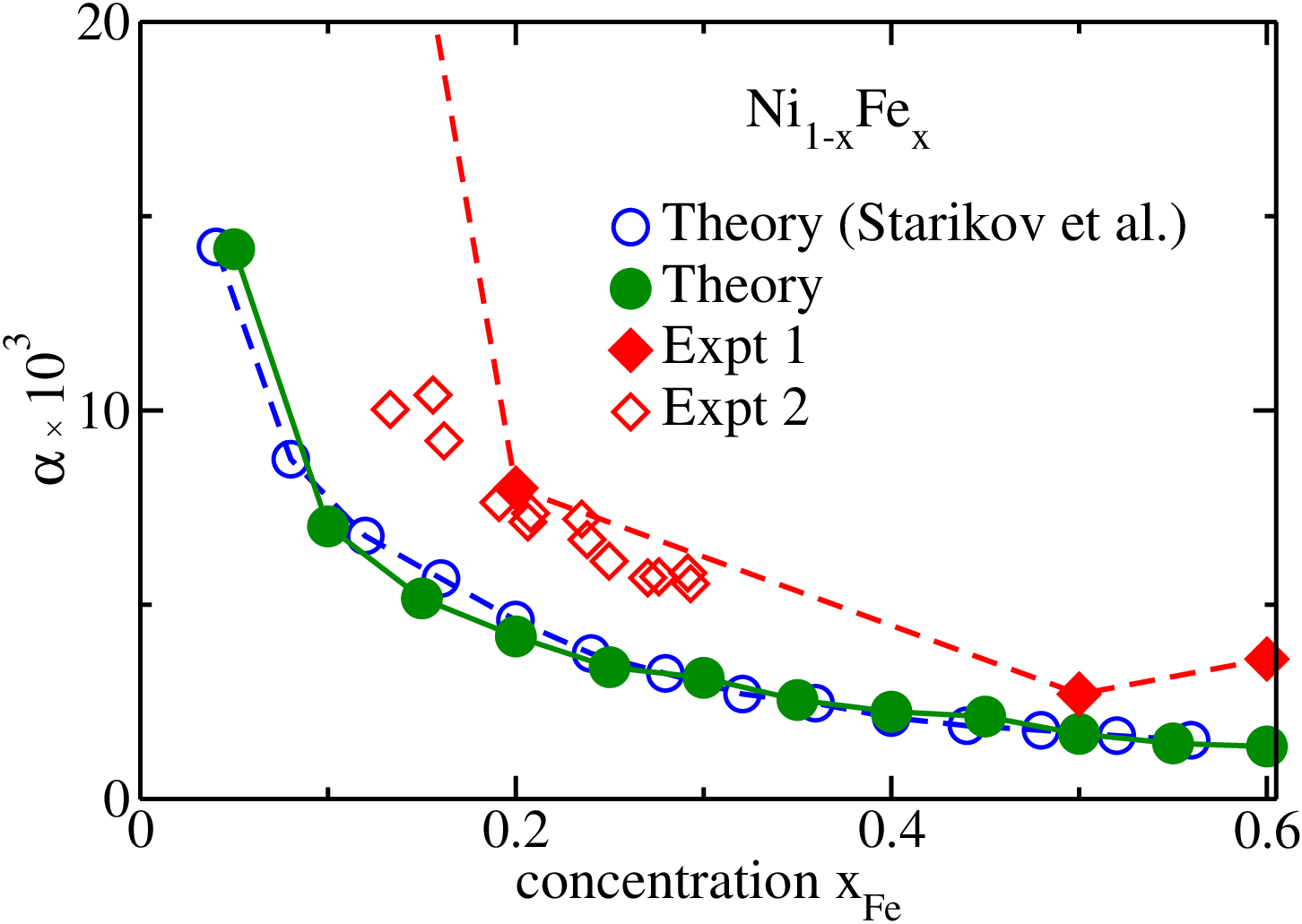}\;(c)
 \includegraphics[scale=0.30]{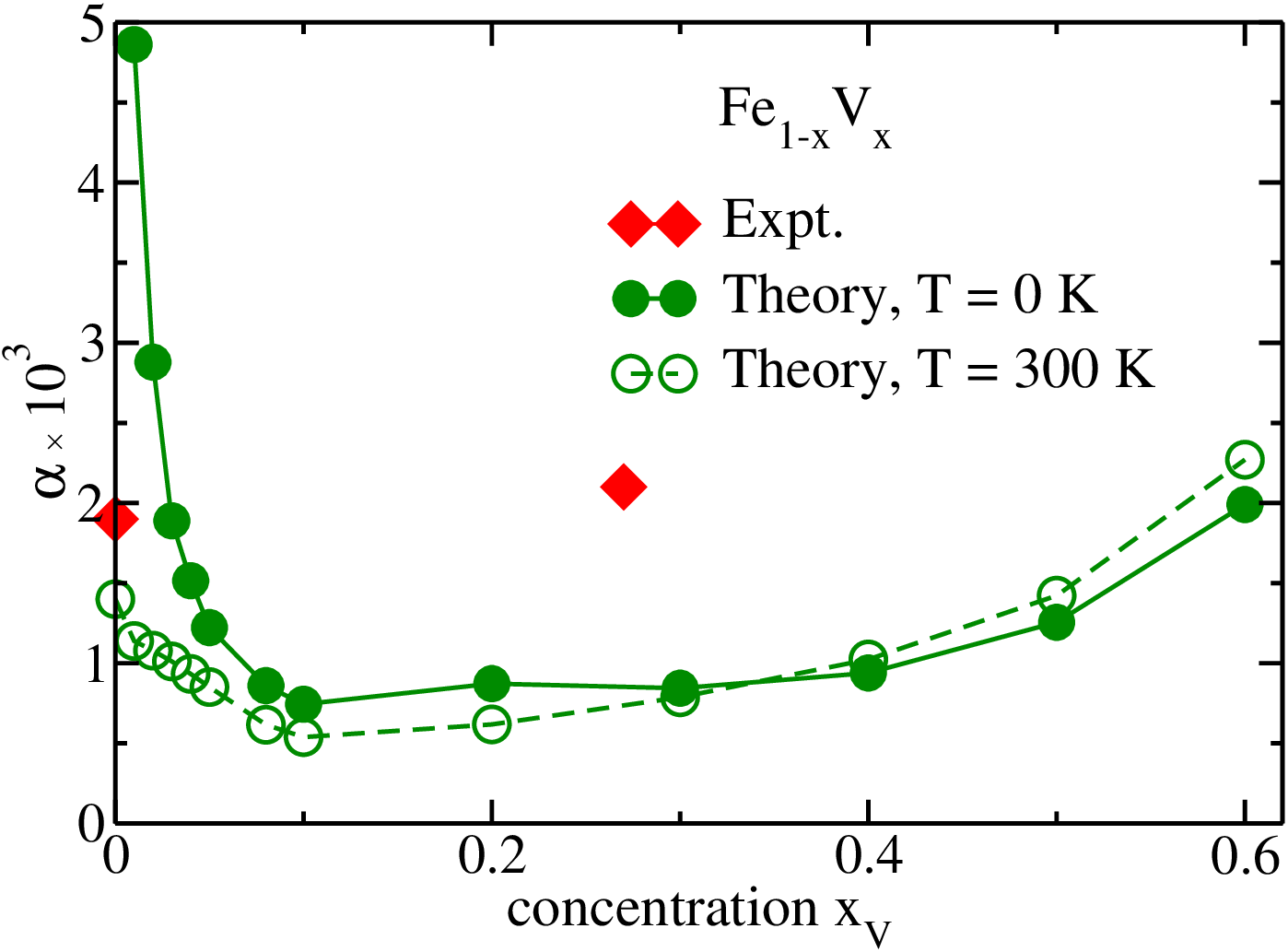}\;(d)
  \end{center}
  \vskip-5mm
  \caption{\label{Fig-FeCoV}The
    Gilbert damping parameter for Fe$_{1-x}$Co$_x$ (a)
    Ni$_{1-x}$Co$_x$ (b) and Ni$_{1-x}$Fe$_x$ (c) as a function 
    of Co and  Fe concentration, respectively:  present
    results within CPA (full circles),
    experimental data by Oogane \cite{OWY+06} (full diamonds).
   (d) Results for bcc Fe$_{1-x}$V$_x$ as a function
    of V concentration:  $T = 0$ K (full circles) and   $T = 300$ K
    (open circles). Squares: experimental data
    \cite{SCB+07}. Open circles: theoretical results by
    Starikov \emph{et al.} \cite{SKB+10}.
}
  \vskip-6mm  
\end{figure}

Fig.\ \ref{Fig-FeCo_BSF}(a)  shows as an example results for the Gilbert
damping parameter $ \alpha (x) $ calculated for bcc Fe$_{1-x}$Co$_{x}$ 
for  $T=0 $~K at different conditions. Full circles represent the
results of the single-cite CPA calculations characterizing the random Fe-Co alloy.
These results are compared to those obtained employing the non-local CPA
\cite{RSG03,KERE07} (NL-CPA) assuming 
no short-range order in the system (opened circles).
Dealing in both cases (CPA and NL-CPA), with completely
disordered system, the NL-CPA maps the alloy
problem on that of an impurity cluster embedded in a translational invariant effective medium
determined selfconsistently, thereby accounting for nonlocal
correlations up to the range of the cluster size. The present 
calculations have been performed for the smallest NL-CPA clusters
containing two sites for bcc based system, accounting 
for the short-range order in the first-neighbor shell.
As one can see, this results in a small decrease of the  $ \alpha$ value 
in the region of concentrations around $x = 0.5$ (opened diamonds), that
is in agreement with the results obtained for partially ordered system
(opened square) for  $x = 0.5$.
The latter have been calculated for the
Fe$_{1-x}$Co$_{x}$ alloy having CsCl structure and alloy components randomly
distributed in two sublattices in the following proportions:
(Fe$_{0.9}$Co$_{0.1}$)(Fe$_{0.1}$Co$_{0.9}$).

Because the moments and spin-orbit coupling strength do not differ very
much for Fe and Co, the variation of $\alpha(x)$ should be determined in the
concentrated regime essentially by the electronic structure at the Fermi
energy $E_F$. As Fig. \ref{Fig-FeCo_BSF}(a) shows, there is indeed a close
correlation with the density of states $n(E_F)$ that may be seen 
as a measure for the number of available relaxation channels.
The change of $\alpha(x)$ due to the increase of the Co concentration is
primarily determined by an apparent shift of the Fermi energy also
varying with concentration (Fig. \ref{Fig-FeCo_BSF}(b)). 
The alloy systems considered have the common feature that the
concentration dependence of $\alpha$ is governed by the 
concentration dependent density of states $n(E_F)$.

A comparison of theoretical $\alpha$ values with the experiment for  bcc
Fe$_{1-x}$Co$_{x}$ is shown in Fig. \ref{Fig-FeCoV}(a), demonstrating
satisfying agreement.
In the case of Ni$_{1-x}$Fe$_{x}$ and Ni$_{1-x}$Co$_{x}$ alloys shown in
Fig. \ref{Fig-FeCoV}, (b) and (c), the Gilbert damping decreases monotonously
with the increase of the Fe and Co concentration, in line with 
experimental data. 
At all concentrations the experimental results are
underestimated by theory approximately by a factor of 2.
The calculated damping parameter $ \alpha (x) $ 
is found in very good agreement with the results 
based on the scattering theory approach \cite{SKB+10}
demonstrating  numerically the equivalence of the two approaches. 
An indispensable requirement to achieve this agreement is to
 include the vertex corrections mentioned above. 
As suggested by Eq. (\ref{alpha_MST}) the variation of $\alpha(x)$ with
concentration $x$ may also reflect to some extent the variation 
of the average magnetic moment of the alloy, $\mu_{tot}$.

The peculiarity of the Fe$_{1-x}$V$_x$ alloy when compared to those
 discussed above is that V is a non-magnetic metal and has only an induced
spin magnetic moment. Despite that, the concentration dependence
of the Gilbert damping parameter at $T = 0$~K for small amounts of V
(see Fig. \ref{Fig-FeCoV}(d)) displays the
same trend as the previously discussed alloys shown in Fig.\ \ref{Fig-FeCoV}(a)-(c).
Taking into account a finite  temperature of $T =
300$~K changes $\alpha$ value  significantly at small V concentrations
leading to an improved agreement with experiment for pure
Fe, while it still compares poorly with the experimental data at $x_V = 0.27$.
One should stress once more that the concentration dependent behavior of
the Gilbert damping parameter of the alloys discussed above is different for
an increased amount of impurities (more than $10\%$), as a result
of a different variation of the DOS $n(E_F)$ caused by a concentration
dependent modification of the electronic states and shift of the Fermi
level.

\subsection{\label{sec:impurities} 5$d$ impurities in 3$d$ transition metals}

\begin{figure}
  \begin{center}
 \includegraphics[scale=0.35]{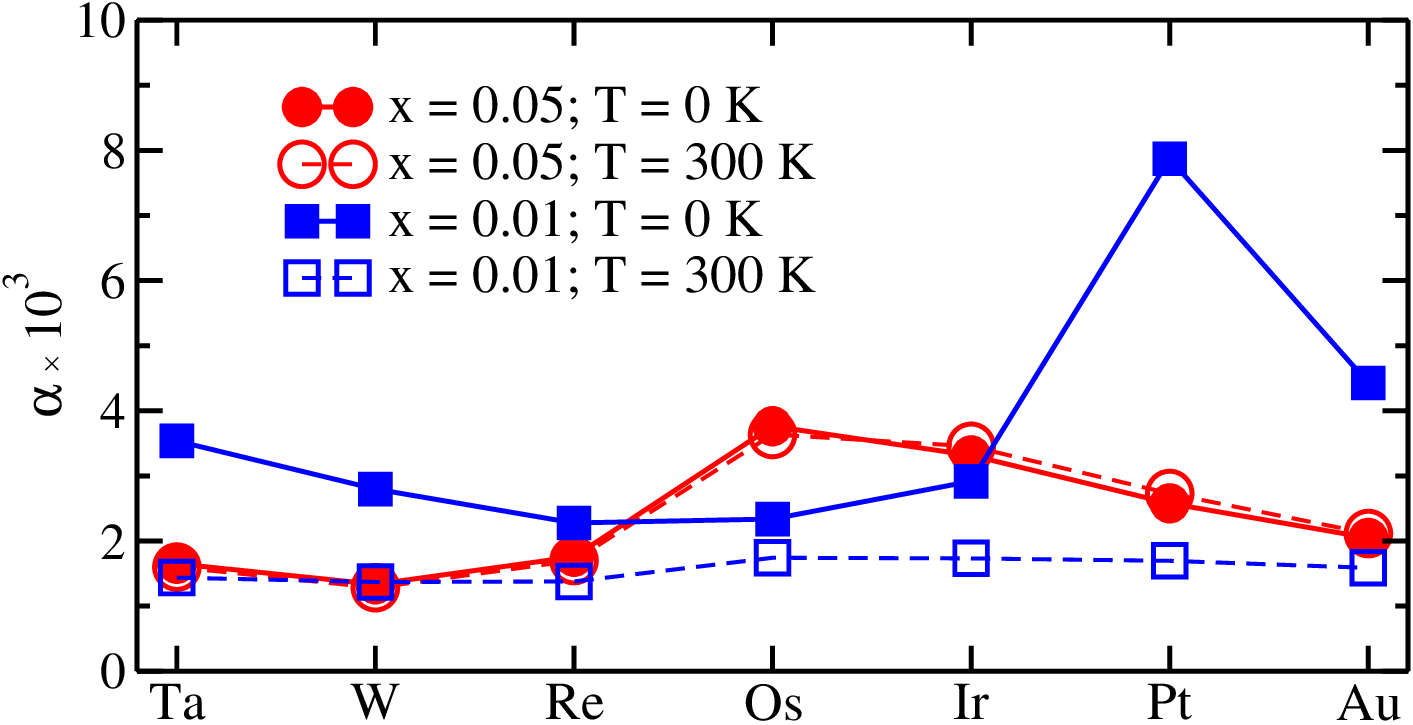}
  \end{center}
  \caption{\label{Fig-Fe5D}
Gilbert damping parameter for bcc Fe with $1\%$ (squares) and $5\%$
(circles) of $5d$ impurities calculated for $T = 0$K (full symbols) and
for $T = 300$K (opened sysmbols).}
\end{figure}

As discussed in our recent work \cite{EMKK11} investigating the temperature
dependent Gilbert damping parameter for pure Ni and  for Ni with Cu
impurities, $\alpha$ is primarily determined by the thermal displacement
in the regime of small impurity concentrations.
This behavior  can also be seen in
Fig. \ref{Fig-Fe5D}, where the results for Fe with
5$d$-impurities are shown. Solid lines
represent results for $T = 0$~K for an impurity
concentration of $1\%$ (full squares) and $5\%$ (full circles). 
As one can see, at smaller concentrations the maximum of the 
Gilbert damping
parameter occurs for Pt. With increasing
impurity content the  $\alpha$ parameter decreases in such a way that
at the concentration of $5\%$ a maximum is observed for Os.

\begin{figure}
  \begin{center}
 \includegraphics[scale=0.35]{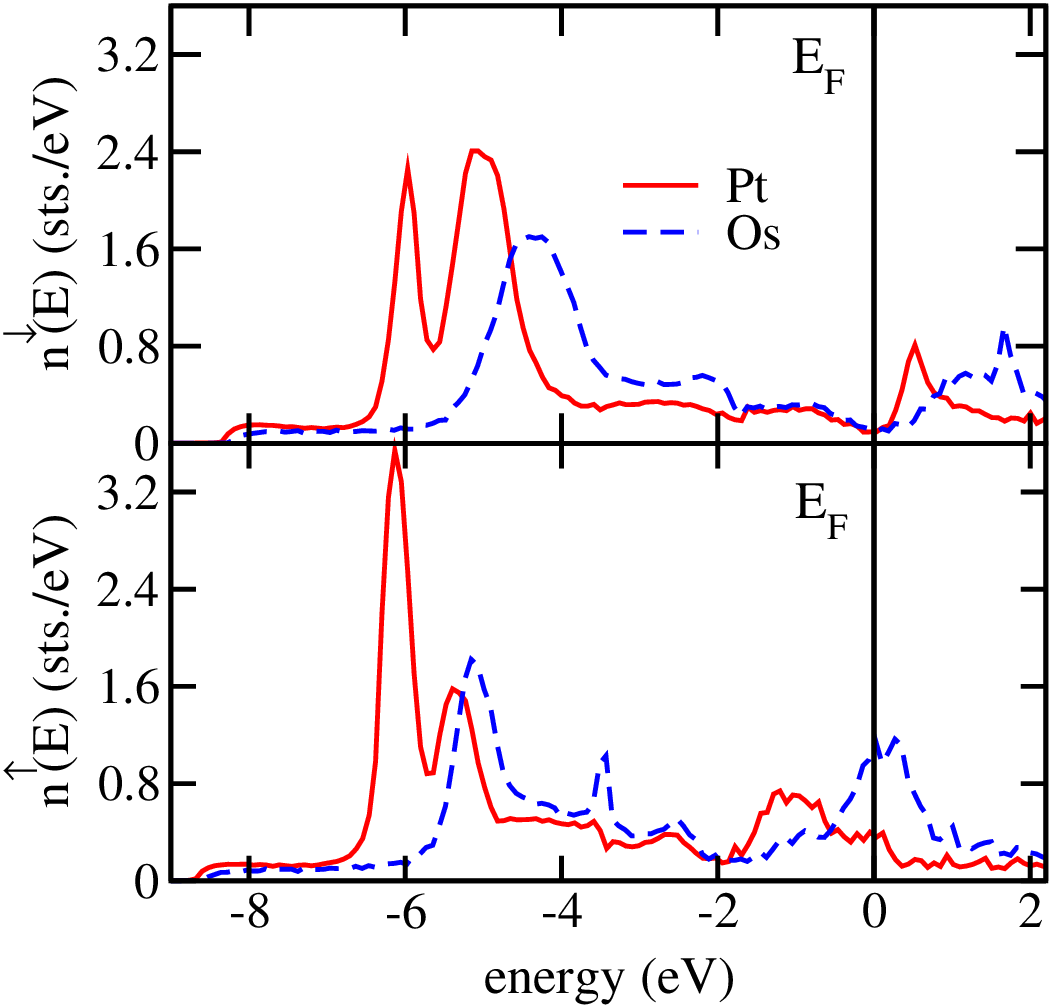} 
  \end{center}
  \caption{\label{Fig-DOS_PtOs}
DOS for Pt in Fe$_{1-x}$Pt$_x$ (full line) and Os in Fe$_{1-x}$Os$_x$ (dashed line)
for $x=0.01$.} 
\end{figure}

The reason for this behavior lies in the rather weak scattering
efficiency of Pt atoms due to a small DOS $n(E_F)$ of the Pt states
when compared for example for Os impurities (see Fig. \ref{Fig-DOS_PtOs}).
This results in a slow decrease of $\alpha$ at
small Pt concentration when the BFS mechanism is mostly responsible for
the energy dissipation. A consequence of this feature can be seen in
the temperature dependence of $\alpha$ ($T = 300$ K, opened squares): a most
pronounced temperature induced decrease of the $\alpha$ value is
observed for Pt and Au.
When the concentration of 5$d$-impurities is increased up to $5\%$, the
 maximum in $\alpha$ occurs for the element with the most efficient
scattering potential 
resulting in
spin-flip scattering processes responsible for dissipation. The
temperature effect at this concentration is very small.

\begin{figure}
  \begin{center}
 \includegraphics[scale=0.35]{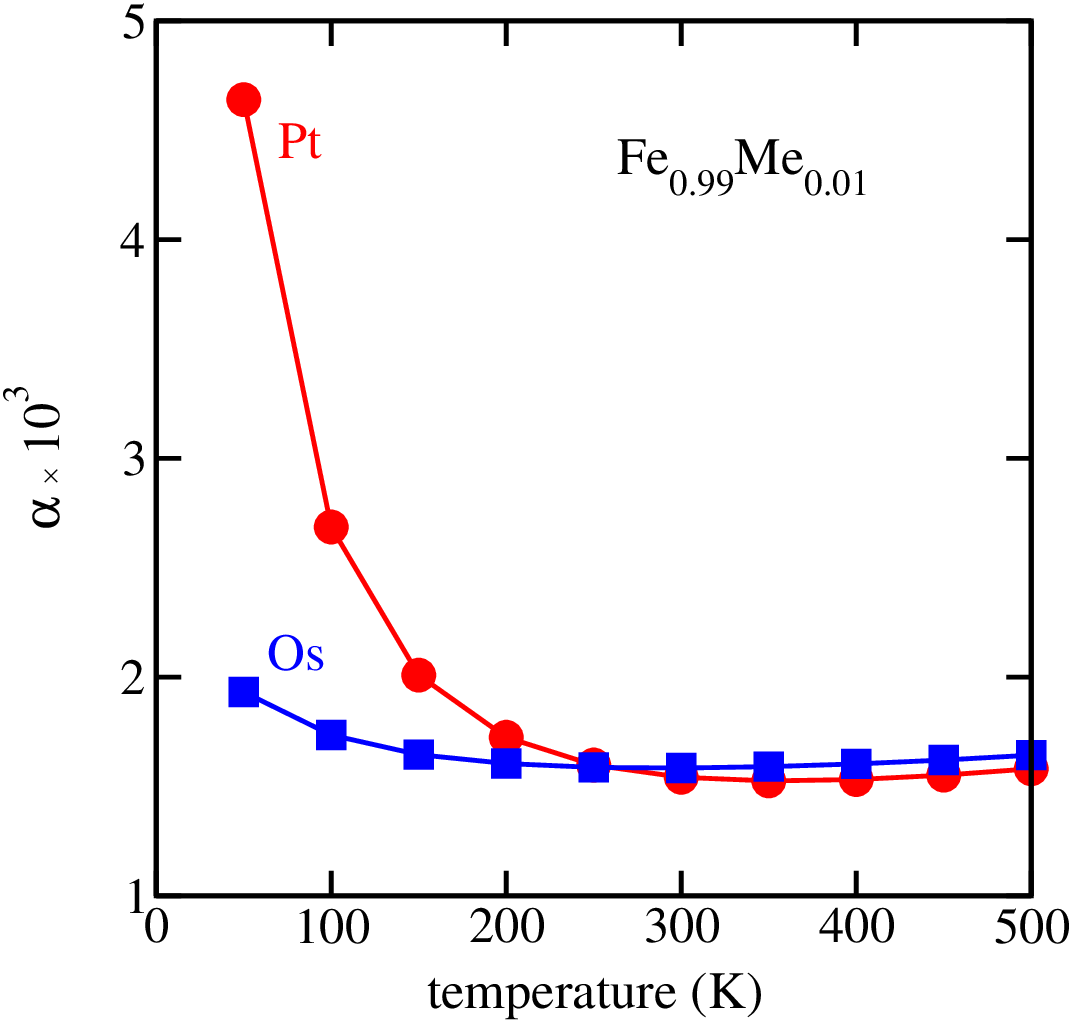} \;(a)
 \includegraphics[scale=0.35]{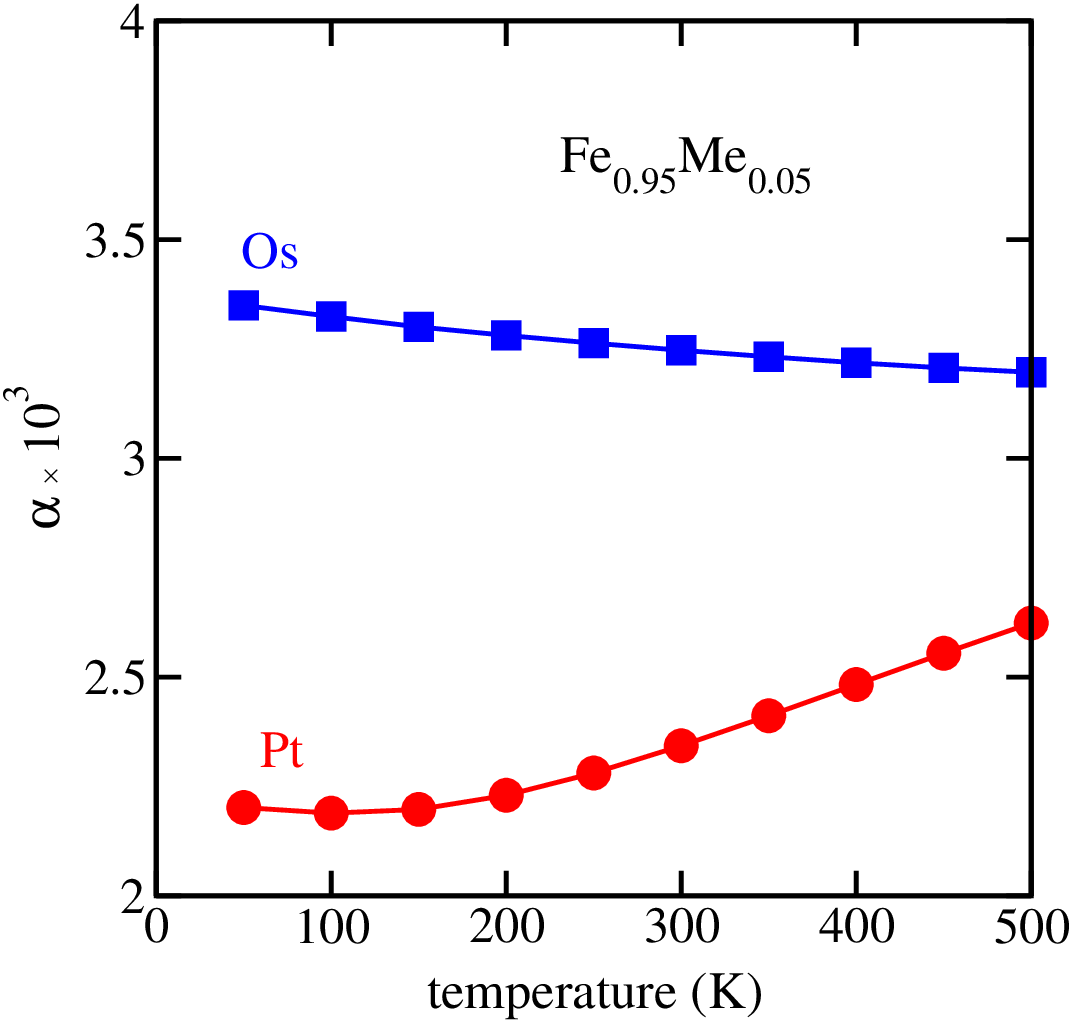} \;(b)
  \end{center}
  \caption{\label{Fig-FePtOs}
Gilbert damping parameter for bcc Fe$_{1-x}$$M_{x}$ with $M =$ Pt (circles) and $M =$ Os (squares)
impurities as a function of temperature for  1\% (a)  and 5\% (b) of the
impurities.} 
\end{figure}

Considering in more detail the temperature dependent behavior of the
Gilbert damping parameter for Fe with Os and Pt impurities, shown in
Fig. \ref{Fig-FePtOs}, one can also observe the consequence of the features 
mentioned above. At $1\%$ of Pt impurities $\alpha$ decreases much
steeper upon increasing the temperature, as compared to the case of Os
impurities. Therefore, in the first case the role of the
scattering processes due to  atomic displacements is much more 
pronounced than in the second case with rather strong scattering
on the Os impurities. When the concentration increases to $5\%$
the dependence of $\alpha$ on the temperature in both cases becomes 
less pronounced.

The previous results can be compared to the results for the 
$5d$-impurities
in the permalloy Fe$_{80}$Ni$_{20}$ (Py), which has been
investigated also experimentally  \cite{RMC+07}.
This system shows some difference in the concentration dependence
when compared with  pure Fe, because Py is a
disordered alloy with a finite value of the $\alpha$ parameter.
Therefore, a substitution of $5d$ impurities leads to a  nearly linear increase of
the Gilbert damping with  impurity content, just as seen in
experiment  \cite{RMC+07}.
%

\begin{figure} 
  \begin{center}
 \includegraphics[scale=0.35]{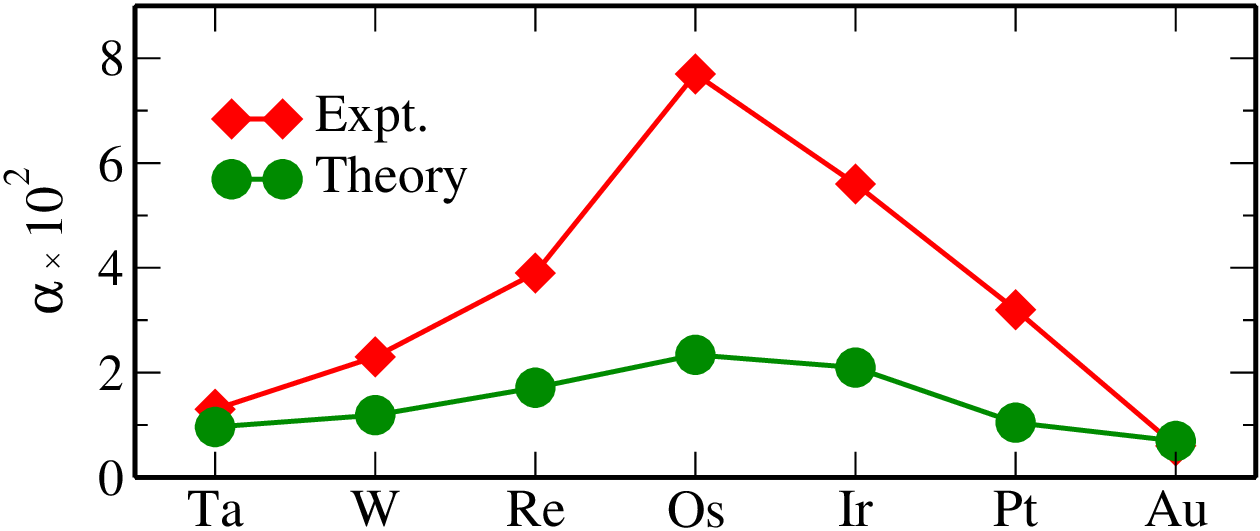}\;(a)
 \includegraphics[scale=0.35]{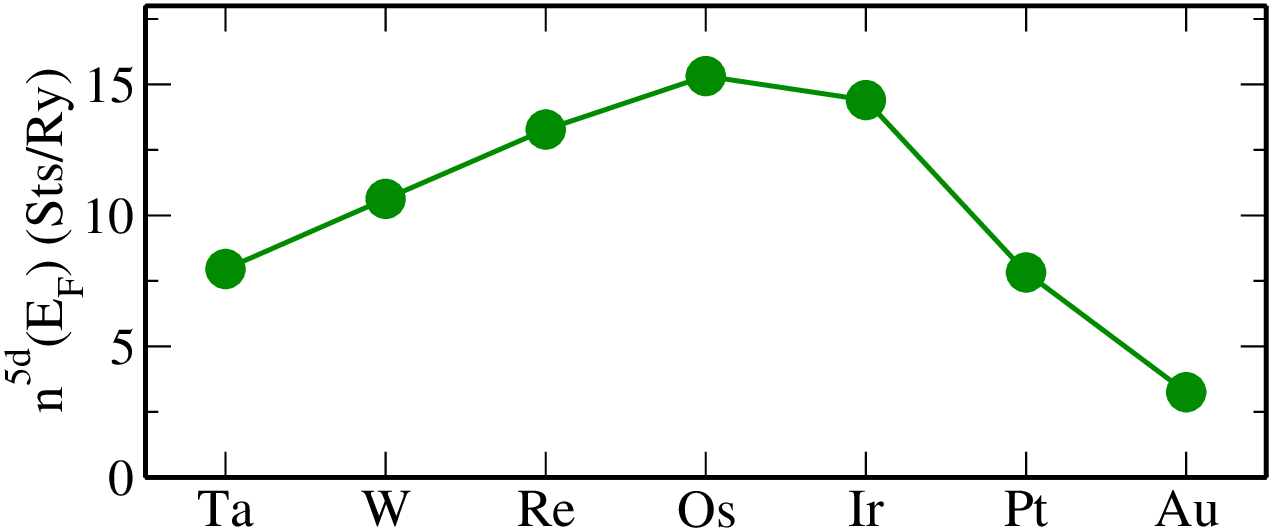}\;(b)
   \end{center}
  \caption{\label{Fig-Py5d}
(a) Gilbert damping parameter $\alpha$ for Py/5d TM systems 
   with 10 \% 5d TM  content in comparison
with  experiment  \cite{RMC+07}; 
(b) spin magnetic moment $m_{spin}^{5d}$ and density of states $n(E_F)$
at the Fermi energy of the $5d$ component in Py/5d TM systems 
   with 10 \% 5d TM content.  }
  \end{figure}

The total damping for $10 \%$ of $5d$-impurities shown in 
Fig.\ \ref{Fig-Py5d}(a) varies roughly parabolically
 over the 5$d$ TM series. 
This variation of $\alpha$ with the type of impurity correlates well
with the density of states $n^{5d}(E_F)$ (Fig. \ref{Fig-Py5d}(b)).
Again the trend of the experimental data is well reproduced by the
calculated values that are however somewhat too low.

\section{\label{sec:metall} Summary}

 In summary, a  formulation for the Gilbert damping parameter
 $ \alpha $ in terms of linear response theory 
 was derived that  led to a  Kubo-Greenwood-like equation. 
 The scheme was implemented using the fully relativistic 
KKR band structure method in combination with the CPA 
alloy theory. 
This allows  to account for various types of scattering mechanisms in 
a parameter-free way, that might be
either due to chemical disorder or to temperature-induced structural
disorder (i.e. electron-phonon scattering effect). The latter has been described by using the
so-called alloy-analogy model  with the 
 thermal displacement of atoms dealt with in a quasi-static manner.
Corresponding applications to pure metals (Fe, Co, Ni) as well as  
to disordered transition-metal alloys led to very good agreement with
results based on the scattering theory approach of Brataas \emph{et al.}\ \cite{BTB08}
and well reproduces the experimental results.
The crucial role of vertex corrections for the Gilbert damping
is demonstrated both  in the case of chemical as well as structural
disorder and the accuracy of finite-temperature results is analyzed 
via test calculations.

Furthermore, the flexibility and numerical efficiency of the present
 scheme was demonstrated by a study on metallic systems on a series
 of binary $3d$-alloys (Fe$_{1-x}$Co$_{x}$,  Ni$_{1-x}$Fe$_x$, 
Ni$_{1-x}$Co$_x$ and Fe$_{1-x}$V$_{x}$), $3d-5d$ TM systems,
 the permalloy-5$d$ TM systems.
The agreement between the present 
 theoretical and experimental results is quite satisfying, although one has to
stress a systematic underestimation of the Gilbert damping by the numerical 
results. This disagreement could be caused either
by the idealized system considered theoretically (e.g., the boundary
effects are not accounted for in present calculations) or because 
of additional intrinsic dissipation mechanisms for bulk systems which
have to be taken into account. These could be, for instance, effects of
temperature induced spin disorder \cite{LSYK11}.




\begin{acknowledgments}

The authors would like to thank the DFG for 
financial support within
the SFB 689 ``Spinph\"anomene in reduzierten Dimensionen''  and within
project EBE-154/23 for financial support.

\end{acknowledgments}




%

\end{document}